# Formation, Habitability, and Detection of Extrasolar Moons

René Heller[1], Darren Williams[2], David Kipping[3], Mary Anne Limbach[4,5], Edwin Turner[4,6], Richard Greenberg[7], Takanori Sasaki[8], Émeline Bolmont[9,10], Olivier Grasset[11], Karen Lewis[12], Rory Barnes[13,14], Jorge I. Zuluaga[15]

## Abstract

The diversity and quantity of moons in the Solar System suggest a manifold population of natural satellites to exist around extrasolar planets. Of peculiar interest from an astrobiological perspective, the number of sizable moons in the stellar habitable zones may outnumber planets in these circumstellar regions. With technological and theoretical methods now allowing for the detection of sub-Earth-sized extrasolar planets, the first detection of an extrasolar moon appears feasible. In this review, we summarize formation channels of massive exomoons that are potentially detectable with current or near-future instruments. We discuss the orbital effects that govern exomoon evolution, we present a framework to characterize an exomoon's stellar plus planetary illumination as well as its tidal heating, and we address the techniques that have been proposed to search for exomoons. Most notably, we show that natural satellites in the range of 0.1 to 0.5 Earth mass (i.) are potentially habitable, (ii) can form within the circumplanetary debris and gas disk or via capture from a binary, and (iii.) are detectable with current technology.

## 1. Introduction

Driven by the first detection of an extrasolar planet orbiting a Sun-like star almost 20 years ago (Mayor and Queloz, 1995), the search for these so-called exoplanets has nowadays achieved the detection of over one thousand such objects and several thousand additional exoplanet candidates (Batalha et al., 2013). Beyond revolutionizing mankind's understanding of the formation and evolution of planetary systems, these discoveries allowed scientists a first approach towards detecting habitats outside the Solar System. Ever smaller and lighter exoplanets were found around Sun-like stars, the record holder now smaller than Mercury (Barclay et al., 2013). Furthermore, ever longer orbital periods can be traced, now encompassing Earth-sized planets in the stellar habitable zones (Quintana et al. 2014) and beyond. While the realm of extrasolar planets is being explored in ever more detail, a new class of objects may soon become accessible to observations: extrasolar moons. These are the natural satellites of exoplanets, and based on our knowledge from the Solar System planets, they may be even more abundant.

    Moons are tracers of planet formation, and as such their discovery around extrasolar planets would fundamentally reshape our understanding of the formation of planetary systems. As an example, the most massive planet in the Solar System, Jupiter, has four massive moons – named Io, Europa, Ganymede, and Callisto – whereas the second-most massive planet around the Sun, Saturn, hosts only one major moon, Titan. These different architectures were likely caused by different termination time scales of gas infall onto the circumplanetary disks, and they show evidence that Jupiter was massive enough to open up a gap in the circumsolar primordial gas and debris disk, while Saturn was not (Sasaki et al., 2010).

    Besides clues to planet formation, exomoons excite the imagination of scientists and the public related to their possibility of being habitats for extrasolar life (Reynolds et al., 1987; Williams et al., 1997; Heller and Barnes, 2013). This idea has its roots

---

[1] Origins Institute, McMaster University, Department of Physics and Astronomy, Hamilton, ON L8S 4M1, Canada; rheller@physics.mcmaster.ca
[2] Penn State Erie, The Behrend College School of Science, 4205 College Drive, Erie, PA 16563-0203; dmw145@psu.edu
[3] Harvard-Smithsonian Center for Astrophysics, 60 Garden Street, Cambridge, MA 02138, USA; dkipping@cfa.harvard.edu
[4] Department of Astrophysical Sciences, Princeton University, Princeton, NJ 08544, USA; mapeters@princeton.edu
[5] Department of Mechanical and Aerospace Engineering, Princeton University, Princeton, NJ 08544, USA
[6] The Kavli Institute for the Physics and Mathematics of the Universe, The University of Tokyo, Kashiwa 227-8568, Japan; elt@astro.princeton.edu
[7] Lunar and Planetary Laboratory, University of Arizona, 1629 East University Blvd, Tucson, AZ 85721-0092, USA; greenberg@lpl.arizona.edu
[8] Department of Astronomy, Kyoto University, Kitashirakawa-Oiwake-cho, Sakyo-ku, Kyoto 606-8502, Japan; takanori@kusastro.kyoto-u.ac.jp
[9] Université de Bordeaux, LAB, UMR 5804, 33270 Floirac, France; emeline.bolmont@obs.u-bordeaux1.fr
[10] CNRS, LAB, UMR 5804, F-33270, Floirac, France
[11] Planetology ad Geodynamics, University of Nantes, CNRS, France; olivier.grasset@univ-nantes.fr
[12] Earth and Planetary Sciences, Tokyo Institute of Technology, Japan; karen.michelle.lewis@gmail.com
[13] Astronomy Department, University of Washington, Box 351580, Seattle, WA 98195, USA; rory@astro.washington.edu
[14] NASA Astrobiology Institute – Virtual Planetary Laboratory Lead Team, USA
[15] FACom - Instituto de Física - FCEN, Universidad de Antioquia, Calle 70 No. 52-21, Medellín, Colombia; jzuluaga@fisica.udea.edu.co

in certain Solar System moons, which may – at least temporarily and locally – provide environments benign for certain organisms found on Earth. Could those niches on the icy moons in the Solar System be inhabited? And in particular, shouldn't there be many more moons outside the Solar System, some of which are not only habitable beyond a frozen surface but have had globally habitable surfaces for billions of years?

While science on extrasolar moons remains theoretical as long as no such world has been found, predictions can be made about their abundance, orbital evolution, habitability, and ultimately their detectability. With the first detection of a moon outside the Solar System on the horizon (Kipping et al., 2012; Heller 2014), this paper summarizes the state of research on this fascinating, upcoming frontier of astronomy and its related fields.

To begin with, we dedicate Section 2 of this paper to the potentially habitable icy moons in the Solar System. This section shall provide the reader with a more haptic understanding of the possibility of moons being habitats. In Section 3, we tackle the formation of natural satellites, thereby focussing on the origin of comparatively massive moons roughly the size of Mars. These moons are suspected to be a bridge between worlds that can be habitable in terms of atmospheric stability and magnetic activity on the one hand, and that can be detectable in the not-too-far future on the other hand. Section 4 is devoted to the orbital evolution of moons, with a focus on the basics of tidal and secular evolution in one or two-satellite systems. This will automatically lead us to tidal heating and its effects on exomoon habitability, which we examine in Section 5, together with aspects of planetary evolution, irradiation effects on moons, and magnetic protection. In Section 6, we outline those techniques that are currently available to search for and characterize exomoons. Section 7 presents a summary and Section 8 an outlook.

## 2. Habitable Niches on Moons in the Solar System

Examination of life on Earth suggests that ecosystems require liquid water, a stable source of energy, and a supply of nutrients. Remarkably, no other planet in the Solar System beyond Earth presently shows niches that combine all of these basics. On Mars, permanent reservoirs of surface water may have existed billions of years ago, but today they are episodic and rare. Yet, we know of at least three moons that contain liquids, heat, and nutrients. These are the Jovian companion Europa, and the Saturnian satellites Enceladus and Titan. Ganymede's intrinsic and induced magnetic dipole fields, suggestive of an internal heat reservoir and a liquid water ocean, make this moon a fourth candidate satellite to host a subsurface habitat. What is more, only four worlds in the Solar System other than Earth are known to show present tectonic or volcanic activity. These four objects are not planets but moons: Jupiter's Io and Europa, Saturn's Enceladus, and Neptune's Triton.

Naturally, when discussing the potential of yet unknown exomoons to host life, we shall begin with an inspection of the Solar System moons and their prospects of being habitats. While the exomoon part of this review is dedicated to the surface habitability of relatively large natural satellites, the following Solar Systems moons are icy worlds that could only be habitable below their frozen surfaces, where liquid water may exist.

### 2.1 Europa

Europa is completely surrounded by a global ocean that contains over twice the liquid water of Earth on a body about the size of Earth's Moon. Its alternative heat source to the weak solar irradiation is tidal friction. Tidal heating tends to turn itself off by circularizing orbits and synchronizing spins. However, Europa's orbit is coupled to the satellites Io and Ganymede through the Laplace resonance (Peale et al., 1979; Greenberg, 1982). The orbital periods are locked in a ratio of 1:2:4, so their mutual interactions maintain eccentricities. As a result, enough heat is generated within Europa to maintain a liquid subsurface ocean about 10 to 100 km deep, which appears to be linked to the surface.

Its surface is marked by dark lines and splotches, and the low rate of impact craters suggests that the surface is younger than 50 Myr (Zahnle et al., 2003). Tectonics produce linear features (cracks, ridges, and bands), and thermal effects produce splotches (chaotic terrain) (Fig. 1). These global scale lineaments roughly correlate with the patterns of expected tidal stress on the ice shell (Helfenstein and Parmentier, 1983; Greenberg et al., 1998) and may record past deviations from uniform synchronous rotation. Double ridges likely form on opposite sides of cracks due to the periodic tidal working over each 80 hr orbit, thereby squeezing up material from the crack and out over the surface. This process would be especially effective if the crack extended from the ocean (Greenberg et al., 1998). Although tidal tension is adequate to crack ice, at depth the overburden pressure counteracts this stress. Therefore, it is unlikely that the ice is thicker than 10 km if ridges form in this way.

The most distinctive crack patterns associated with tidal stress are the cycloids, chains of arcs, each about 100 km long, connected at cusps, and extending often over 1000 km or more. Cycloids are ubiquitous on Europa, and they provided the first observational evidence for a liquid water ocean, because they fit the tidal stress model so well and because adequate tidal distortion of Europa would be impossible if all the water were frozen (Hoppa et al., 1999; Groenleer and Kattenhorn, 2008). Given that Europa's surface age is less than 1% of the age of the Solar System, and that many cycloids are among the freshest features, it seemed evident that the ocean must still exist today. Confirmation came with measurements of Jupiter's magnetic field near the satellite, which showed distortions consistent with the effect of a near-surface electrical conducting salty ocean (Kivelson et al., 2000). Active plumes observed at the moon's south pole deliver further evidence of a liquid water reservoir, but it remains unclear whether these waters that feed these geysers are connected to the global subsurface ocean or they are local (Roth et al., 2014).

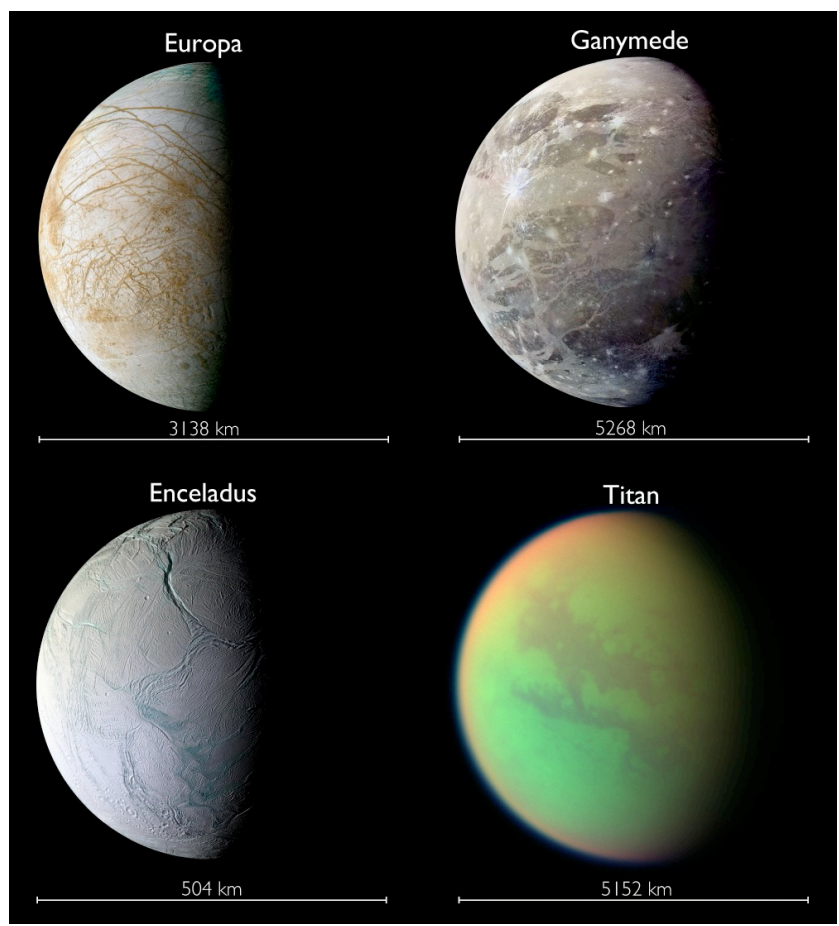

**Figure 1**: Europa, Enceladus, Ganymede, and Titan are regarded as potentially habitable moons. Global lineaments on Europa's surface and ridges on Enceladus indicate liquid water as close as a few kilometers below their frozen surfaces. Ganymede's surface is much older with two predominant terrains, bright, grooved areas and older, heavily cratered, dark regions. Titan has a dense nitrogen atmosphere and liquid methane/ethane seas on its surface. While the atmosphere is intransparent to the human eye, the lower right image contains information taken in the infrared. Note the different scales! Moon diameters are indicated below each satellite. [Image credits: NASA/JPL/Space Science Institute/Ted Stryk]

Large plates of Europa's surface ice have moved relative to one another. Along some cracks, often hundreds of kilometers long, rifts (called dilation bands) have opened up and filled with new striated ice (Tufts et al., 2000). The dilation can be demonstrated by reconstructing the surface, matching opposite sides, like pieces of a picture puzzle. The apparent mobility of large plates of surface ice shows that the cracks must penetrate to a fluid or low-viscosity layer, again indicating that the ice is less than 10 km thick.

Chaotic terrain covers nearly half of Europa's surface (Riley et al., 2000) and appears to have been thermally disrupted, leaving a lumpy matrix between displaced rafts, on whose surfaces fragments of the previous surface are clearly visible. The crust appears to have melted, allowing blocks of surface ice to float, move, and tilt before refreezing back into place (Carr et al., 1998). Only modest, temporary, local or regional concentrations of tidal heat are required for substantial melt-through. If 1% of the total internal heat flux were concentrated over an area about 100 km across, the center of such a region would melt through in only a few thousand years and cause broad exposure over tens of km wide in $10^4$ yr (O'Brien et al., 2000).

If the ice were thick, the best prospects for life would be near the ocean floor, where volcanic vents similar to those on the Earth's ocean floor could support life by the endogenic substances and tidal heat. Without oxygen, organisms would require alternative metabolisms whose abundance would be limited (Gaidos et al., 1999). For a human exploration mission, Europan life would only be accessible after landing on the surface, penetrating about 10 km of viscous ice, and diving 100 km or more to the ocean floor, where the search would begin for a hypothetical volcanic vent. If the ice is thin enough to be permeable, the odds that life is present and detectable increase. At the surface, oxidants – especially oxygen and hydrogen peroxide – are produced by impacts of energetic charged particles trapped in the Jovian magnetic field (Hand et al., 2007). Although this radiation must also sterilize the upper 10 cm of ice, organisms might be safe below that level. Photosynthesis would be possible down to a depth of a few meters, and the oxidants, along with organic compounds from cometary debris, are mixed to that depth by micrometeoroid impacts. Any active crack, periodically opening and closing with the tide, might allow the vertical flow of liquid water from the ocean and back down. Organisms living in the ice or the crack might take advantage of the access to near-surface oxidants and organics as well as oceanic substances and the flow of warm (0°C) water (Greenberg et al., 2002). Based on the size of the double ridges, cracks probably remain active for tens of thousands of years, and so organisms would have time to thrive. But once a crack freezes, they would need to hibernate in the ice or migrate into the ocean or to another active site. Exposure of water at the surface would allow some oxygen to enter the ocean directly. The gradual build-up of frozen ocean water over the surface exposes fresh ice to the production of oxidants and also buries ever deeper the previously oxygenated ice. Based on resurfacing rates of the various geological processes, oxygen may enter the ocean at a rate of about $3 \times 10^{11}$ moles/yr (Greenberg, 2010), equivalent to the respiration requirements of 3 million tons of terrestrial fish. Hence, this delivery rate could allow a concentration of oxygen adequate to support complex life, which is intrinsically less efficient than microorganisms, due to their extra operational overhead. Moreover, with this oxygen source, an ecosystem could be independent of photosynthesis. In the ocean, interaction of the oxygen with the rocky or clay seafloor could be significant and eventually result in a drastic decrease in the pH of the water (Pasek and Greenberg, 2012). How much this process could affect life depends on the efficiency of the contact between the water and the rock. Moreover, if organisms consume the oxygen fast enough, they could ameliorate the acidification.

## 2.2 Ganymede

With its 2634 km in radius, the most massive moon in the Solar System, Ganymede, features old, densely cratered terrain and widespread regions that may have been subject to tectonic resurfacing. The great variety in geologic and geomorphic units has been dated over a range of several billions of years, and shows evidence of past internal heat release. Besides Mercury and Earth, Ganymede is one of only three solid bodies in the Solar System that generate a magnetic dipole field. It also possesses a small moment of inertia factor of 0.3115 (Schubert et al., 2004)[16], which is indicative of a highly differentiated body. This moon is thought to have (i) an iron-rich core, in which a liquid part must be present to generate the magnetic field, (ii) a silicate shell, (iii) a hydrosphere at least 500 km thick, and (iv) a tenuous atmosphere (Anderson et al., 1996; Kivelson et al., 2002; Sohl et al., 2002; Spohn and Schubert, 2003).

There is no evidence of any present geologic activity on Ganymede. Locally restricted depressions, called "paterae", may have formed through cryovolcanic processes, and at least one of them is interpreted as an icy flow that produced smooth bright lanes within the grooved terrain (Giese et al., 1998; Schenk et al., 2004). This suggests that cryovolcanism and tectonic processes played a role in the formation of bright terrain. There is no geologic evidence on Ganymede that supports the existence of shallow liquid reservoirs. And analyses of magnetic field measurements collected by the Galileo space probe are still inconclusive regarding an interpretation of a global subsurface ocean (Kivelson et al., 1996). The challenge is in the complex interaction among four field components, namely, a possibly induced field, the moon's intrinsic field, Jupiter's magnetosphere, and the plasma environment. But still, the presence of an ocean is in agreement with geophysical models, which predict that tidal dissipation and radiogenic energy keep the water liquid (Spohn and Schubert, 2003; Hussmann et al., 2006; Schubert et al. 2010).

While Europa with its relatively thin upper ice crust and a global ocean, which is likely in contact with the rocky ocean floor, has been referred to as a class III habitat (Lammer et al., 2009)[17], Ganymede may be a class IV habitat. In other words, its hydrosphere is split into (i) a high-pressure ice layer at the bottom that consists of various water-rich ices denser than liquid water, (ii) a subsurface water ocean that is likely not in direct contact with the underlying silicate floor, and (iii) an ice-I layer forming the outer crust of the satellite. In this model, Ganymede serves as an archetype for the recently suggested class of water world extrasolar planets (Kuchner, 2003; Léger et al., 2004; Kaltenegger et al., 2013; Levi et al., 2013). Ganymede's liquid layer could be up to 100 km thick (Sohl et al., 2010), and it is not clear whether these deep liquid oceans can be habitable. Chemical and energy exchanges between the rocky layer and the ocean, which are crucial for habitability, cannot be ruled out, but they require efficient transport processes through the thick high-pressure icy layer. Such processes are indeed possible (Sohl et al., 2010) but not as clear-cut as the exchanges envisaged for Europa, where they probably prevailed

---

[16] This dimensionless measure of the moment of inertia equals $I_{Cal}/(M_{Cal} \, R_{Cal}^2)$, where $I_{Cal}$ is Callisto's moment of inertia, $M_{Cal}$ its mass, and $R_{Cal}$ its radius. This factor is 0.4 for a homogeneous spherical body but less if density increases with depth.
[17] See cases 3 and 4 in their Fig. 11.

until recent times.

## 2.3 Enceladus

Another potential moon habitat is Enceladus. With an average radius of merely 250 km, Saturn's sixth-largest moon should be a cooled, dead body, if it were not subject to intense endogenic heating due to tidal friction. Contrary to the case of Europa, Enceladus' orbital eccentricity of roughly 0.0047 cannot be fully explained by gravitational interactions with its companion moons. While perturbations from Dione may play a role, they cannot explain the current thermal flux observed on Enceladus (Meyer and Wisdom, 2007). The current heat flux may actually be a remainder from enhanced heating in the past, and it has been shown that variations in the strength of this heat source likely lead to episodic melting and resurfacing events (Běhounková et al., 2012).

Given the low surface temperatures on Enceladus of roughly 70 K, accompanied by its low surface gravity of 0.114 m/s, viscous relaxation of its craters is strongly inhibited. The existence of a large number of shallow craters, however, suggests that subsurface temperatures are around 120 K or higher and that there should have been short periods of intense heating with rates up to 0.15 W/m$^2$ (Bland et al., 2012). These conditions agree well with those derived from studies of its tectonically active regions, which yield temporary heating of 0.11 to 0.27 W/m$^2$ (Bland et al., 2007; Giese et al., 2008).

Enceladus' internal heat source results in active cryovolcanic geysers on the moon's southern hemisphere (Porco et al., 2006; Hansen et al., 2006). The precise location coincides with a region warm enough to make Enceladus the third body, after Earth and Io, whose geological energy flow is accessible by remote sensing. Observations of the *Cassini* space probe during close flybys of Enceladus revealed temperatures in excess of 180 K along geological formations that have now become famous as "tiger stripes" (Porco et al, 2006; Spencer et al., 2008) (Fig. 1). Its geological heat source is likely strong enough to sustain a permanent subsurface ocean of liquid saltwater (Postberg et al., 2011). In addition to water ($H_2O$), traces of carbon dioxide ($CO_2$), methane ($CH_4$), ammonia ($NH_3$), salt (NaCl), and $^{40}$Ar have been detected in material ejected from Enceladus (Waite et al., 2006; Waite et al., 2009; Hansen et al., 2011). The solid ejecta, about 90% of which fall back onto the moon (Hedman et al, 2009; Postberg et al., 2011), cover its surface with a highly reflective blanket of μm-sized water ice grains. With a reflectivity of about 0.9, this gives Enceladus the highest bond albedo of any body in the Solar System (Howett et al., 2010).

While Europa's subsurface ocean is likely global and may well be in contact with the moon's silicate floor, Enceladus' liquid water reservoir is likely restricted to the thermally active south polar region. It is not clear if these waters are in contact with the rocky core (Tobie et al., 2008; Zolotov, 2007) or if they only form pockets in the satellite's icy shell (Lammer et al., 2009).

All of these geological activities, and in particular their interference with liquid water, naturally open up the question of whether Enceladus may be habitable. Any ecosystem below the moon's frozen ice shield would have to be independent from photosynthesis. It could also not be based on the oxygen ($O_2$) or the organic compounds produced by surface photosynthesis (McKay et al., 2008). On Earth, indeed, such subsurface ecosystems exist. Two of them rely on methane-producing microorganisms (methanogens), which themselves feed on molecular hydrogen ($H_2$) released by chemical reactions between water and olivine rock (Stevens and McKinley, 1995; Chapelle et al., 2002). A third such anaerobic ecosystem is based on sulfur-reducing bacteria (Lin et al, 2006). The $H_2$ required by these communities is ultimately produced by the radioactive decay of long-lived uranium (U), thorium (Th), and potassium (K). It is uncertain whether such ecosystems could thrive on Enceladus, in particular due to the possible lack of redox pairs under a sealed ocean (Gaidos et al., 1999). For further discussion of possible habitats on Enceladus, see the work of McKay et al. (2008).

What makes Enceladus a particularly interesting object for the in-situ search for life is the possibility of a sample return mission that would not have to land on the moon's surface (Reh et al., 2007; Razzaghi et al., 2007; Tsou et al., 2012). Instead, a spacecraft could repeatedly dive through the Enceladian plumes and collect material that has been ejected from the subsurface liquid water reservoir. This icy moon may thus offer a much more convenient and cheap option than Mars from which to obtain biorelevant, extraterrestrial material. What is more, once arrived in the Saturnian satellite system, a spacecraft could even take samples of the upper atmosphere of Titan (Tsou et al., 2012).

## 2.4 Titan

With its nitrogen-dominated atmosphere and surface pressures of roughly 1.5 bar, Titan is the only world in the Solar System to maintain a gaseous envelope at least roughly similar to that of Earth in terms of composition and pressure. It is also the only moon beyond Earth's Moon and the only object farther away from the Sun than Mars, from which a spacecraft has returned in-situ surface images. Footage sent back from the surface by the Huygens lander in January 2005 show pebble-sized, rock-like objects some ten centimeters across (mostly made of water and hydrocarbon ices) on a frozen ground with compression properties similar to wet clay or dry sand (Zarnecki et al., 2005). Surface temperatures around 94 K imply that water is frozen and cannot possibly play a key role in the weather cycle as it does on Earth. Titan's major atmospheric constituents are molecular nitrogen ($N_2$, 98.4 %), $CH_4$ (1.4 %), molecular hydrogen (0.1 %), and smaller traces of acetylene ($C_2H_2$) as well as ethane ($C_2H_6$) (Cousteinis et al., 2007). The surface is hidden to the human eye under an optically thick photochemical haze (Fig. 1), which is composed of various hydrocarbons. While the surface illumination is extremely low, atmospheric $C_2H_2$ could act as a mediator and transport the energy of solar ultra-violet radiation and high-energy particles to the surface, where it could undergo exothermic reactions (Lunine, 2010). Intriguingly, methane should be irreversibly destroyed by photochemical processes on a timescale of 10 to 100 million years (Yung et al., 1984; Atreya et al., 2006). Hence, its abundance suggests that it is continuously resupplied. Although possibilities of methanogenic life on Titan have been hypothesized (McKay and Smith, 2005; Schulze-Makuch and Grinspoon, 2005), a biological origin of methane seems unlikely because its $^{12}C/^{13}C$ ratio is not enhanced with respect to the Pee Dee Belemnite inorganic standard value (Niemann et al., 2005). Instead, episodic cryovolcanic activity could release substantial amounts of methane from Titan's crust (LeCorre et al., 2008), possibly driven by outgassing from internal reservoirs of clathrate hydrates (Tobie et al., 2006).

In addition to its thick atmosphere, Titan's substantial reservoirs of liquids on its surface make it attractive from an astrobiological perspective (Stofan et al., 2007; Hayes et al., 2011). These ponds are mostly made of liquid ethane (Brown et al., 2008) and they feed a weather cycle with evaporation of surface liquids, condensation into clouds, and precipitation (Griffith et al., 2000). Ultimately, the moon's non-synchronous rotation with respect to Saturn as well as its substantial orbital eccentricity of 0.0288 (Sohl et al., 1995; Tobie et al., 2005) point towards the presence of an internal ocean, the composition and depth of which is unknown (Lorenz et al., 2008; Norman, 2011). If surface life on Titan were to use non-aqueous solvents, where chemical reactions typically occur with much higher rates than in solid or gas phases, it would have to rely mostly on ethane. However, laboratory tests revealed a low solubility of organic material in ethane and other non-polar solvents (McKay, 1996), suggesting also a low solubility in liquid methane. While others have argued that life in the extremely cold hydrocarbon seas on Titan might still be possible (Benner et al., 2004), the characterization of such ecosystems on extrasolar moons will not be possible for the foreseeable future. In what follows, we thus exclusively refer to habitats based on liquid water.

## 3. Formation of Moons

Ganymede, the most massive moon in the Solar System, has a mass roughly 1/40 that of Earth or 1/4 that of Mars. Supposing Jupiter and its satellite system would orbit the Sun at a distance of one astronomical unit (AU), Ganymede's ices would melt but it would not be habitable since its gravity is too small to sustain a substantial atmosphere. Terrestrial sized objects must be larger than 1 to 2 Mars masses, or 0.1 to 0.2 Earth masses ($M_⊕$), to have a long-lived atmosphere and moist surface conditions in the stellar habitable zones (Williams et al., 1997). Hence, to assess the potential of extrasolar moons to have habitable surfaces, we shall consult satellite formation theories and test their predictions for the formation of Mars- to Earth-sized moons.

Ganymede as well as Titan presumably formed in accretion disks surrounding a young Jupiter and Saturn (Canup and Ward, 2002), and apparently there wasn't enough material or accretion efficiency in the disks to form anything larger. Disks around similar giant planets can form much heavier moons if the disk has a lower gas-to-solid ratio or a high viscosity parameter $\alpha$ (Canup and Ward, 2006). Assuming disks similar to those that existed around Jupiter and Saturn, giant exoplanets with masses five to ten times that of Jupiter might accrete moons as heavy as Mars.

The orbit of Triton, the principal moon of Neptune, is strongly suggestive of a formation via capture rather than agglomeration of solids and ices in the early circumplanetary debris disk. The satellite's orbital motion is tilted by about 156° against the equator of its host planet, and its almost perfectly circular orbit is embraced by various smaller moons, some of which orbit Neptune in a prograde sense (that is, in the same direction as the planet rotates) and others which have retrograde orbits. With Triton being the seventh-largest moon in the Solar System and with Earth- to Neptune-sized planets making up for the bulk part of extrasolar planet discoveries (Batalha et al. 2013; Rowe et al. 2014), the capture of substantial objects into stable satellites provides another reasonable formation channel for habitable exomoons.

Moving out from the Solar System towards moons orbiting extrasolar planets, and eventually those orbiting planets in the stellar habitable zones, we ask ourselves: "How common are massive exomoon systems around extrasolar planets?" Ultimately, we want to know the frequency of massive moons the size of Mars or even Earth around those planets. In the following, we discuss recent progress towards addressing these questions from the formation theory point of view.

### 3.1 In-situ formation

### 3.1.1 Planetary disk models

Various models for satellite formation in circumplanetary protosatellite disks have been proposed recently. One is called the solids enhanced minimum mass model (Mosqueira and Estrada, 2003a, 2003b; Estrada et al., 2009), another one is an actively supplied gaseous accretion disk model (Canup and Ward, 2002, 2006, 2009), and a third one relies on the viscous spreading of a massive disk inside the planet's Roche limit (Crida and Charnoz, 2012). All these models have been applied to study the formation of the Jovian and Saturnian satellite systems. In the Mosqueira and Estrada model, satellite formation occurs once sufficient gas has been removed from an initially massive subnebula and turbulence in the circumplanetary disk subsides. Satellites form from solid materials supplied by ablation and capture of planetesimal fragments passing through the massive disk. By contrast, in the Canup and Ward picture satellites form in the accretion disk at the very end of the host planet's own accretion, which should reflect the final stages of growth of the host planets. Finally, the Crida and Charnoz theory assumes that satellites originate at the planet's Roche radius and then move outward due to gravitational torques experienced within the tidal disk. While the latter model may explain some aspects of satellite formation around a wide range of central bodies, that is, from bodies as light as Pluto to giant planets as heavy as Saturn, it has problems reproducing the Jovian moon system.

The actively supplied gaseous accretion disk model postulates a low-mass, viscously evolving protosatellite disk with a peak surface density near 100 g/cm$^2$ that is continuously supplied by mass infall from the circumstellar protoplanetary disk. The temperature profile of the circumplanetary disk is dominated by viscous heating in the disk, while the luminosity of the giant planet plays a minor role. "Satellitesimals", that is, proto-moons are assumed to form immediately from dust grains that are supplied by gas infall. Once a satellite embryo has grown massive enough, it may be pushed into the planet through type I migration driven by satellite-disk interaction (Tanaka et al., 2002). The average total mass of all satellites resulting from a balance between disposal by type I migration and repeated satellitesimal accretion is universally of the order of $10^{-4}$ $M_p$, where $M_p$ is the host planet mass (Canup and Ward, 2006). Beyond a pure mass scaling, Sasaki et al. (2010) were able to explain the different architectures of the Jovian and the Saturnian satellite systems by including an inner cavity in the circum-Jovian disk, while Saturn's disk was assumed to open no cavity, and by considering Jupiter's gap opening within the solar disk. Ogihara and Ida (2012) improved this model by tracing the orbital evolution of moonlets embedded in the circumplanetary disk with an N-body method, thereby gaining deeper insights into the compositional evolution of moons.

### 3.1.2 Differences in the Jovian and Saturnian satellite systems

As a gas giant grows, its gravitational perturbations oppose viscous diffusion and pressure gradients in the gas disk and thereby open up a gap in the circumstellar protoplanetary disk (Lin and Papaloizou, 1985). Since the critical planetary mass for the gap opening at 5 - 10 AU is comparable to Jupiter's mass, it is proposed that Jupiter's final mass is actually determined by the gap opening (Ida and Lin, 2004). While the theoretically predicted critical mass for gap opening generally increases with orbital radius, Saturn has a mass less than one-third of Jupiter's mass in spite of its larger orbital radius. It is conventionally thought that Saturn did not open up a clear gap and the infall rate of mass onto Saturn and its protosatellite disk decayed according to the global depletion of the protoplanetary disk.[18] Since core accretion is generally slower at larger orbital radii in the protoplanetary disk (Lin and Papaloizou, 1985), it is reasonable to assume that gas accretion onto Saturn proceeded in the dissipating protoplanetary disk, while Jupiter's growth was truncated by the formation of a clear gap before the disk lost most of its gas.

Although gap opening may have failed to truncate the incident gas completely, it would certainly reduce the infall rate by orders of magnitude (D'Angelo et al., 2003). After this truncation, the protosatellite disk would be quickly depleted on its own viscous diffusion timescale of about $(10^{-3}/\alpha)10^3$ years, where $\alpha$ represents the strength of turbulence (Shakura and Sunyaev, 1973) and has typical values of $10^{-3}$ to $10^{-2}$ for turbulence induced by magneto-rotational instability (Sano et al., 2004). On the other hand, if Saturn did not open a gap, the Saturnian protosatellite disk was gradually depleted on the much longer viscous diffusion timescales of the protoplanetary disk, which are observationally inferred to be $10^6$ to $10^7$ years (Meyer et al., 2007). The different timescales come from the fact that the protosatellite disk is many orders of magnitude smaller than the protoplanetary disk. The difference would significantly affect the final configuration of satellite systems, because type I migration timescales of protosatellites (approximately $10^5$ years) are in between the two viscous diffusion timescales (Tanaka et al., 2002). Jovian satellites may retain their orbital configuration frozen in a phase of the protosatellite disk with relatively high mass at the time of abrupt disk depletion, while Saturnian satellites must be survivors against type I migration in the final less massive disk.

The different disk masses reflected by the two satellite systems indicate that their disks also had different inner edges. By analogy with the observationally inferred evolution of the inner edges of disks around young stars (Herbst and Mundt, 2005), Sasaki et al. (2010) assumed (i.) a truncated boundary with inner cavity for the Jovian system and (ii.) a non-truncated boundary without cavity for the Saturnian system. Condition (i.) may be appropriate in early stages of disk evolution with high disk mass and accretion rate and strong magnetic field, while in late stages with a reduced disk mass, condition (ii.) may be more adequate. The Jovian satellite system may have been frozen in a phase with condition (i.), since the depletion timescale of its protosatellite disk is much shorter than typical type I migration timescales of protosatellites. In this system, type I migration of satellites is terminated by a local pressure maximum near the disk edge. On the other hand, the Saturnian system that formed in a gradually dissipating protosatellite disk may reflect a later evolution phase with condition II.

---

[18] The recently proposed Grand Tack model (Walsh et al., 2011), however, suggests that Jupiter and Saturn actually opened up a common gap (Pierens and Raymond, 2011).

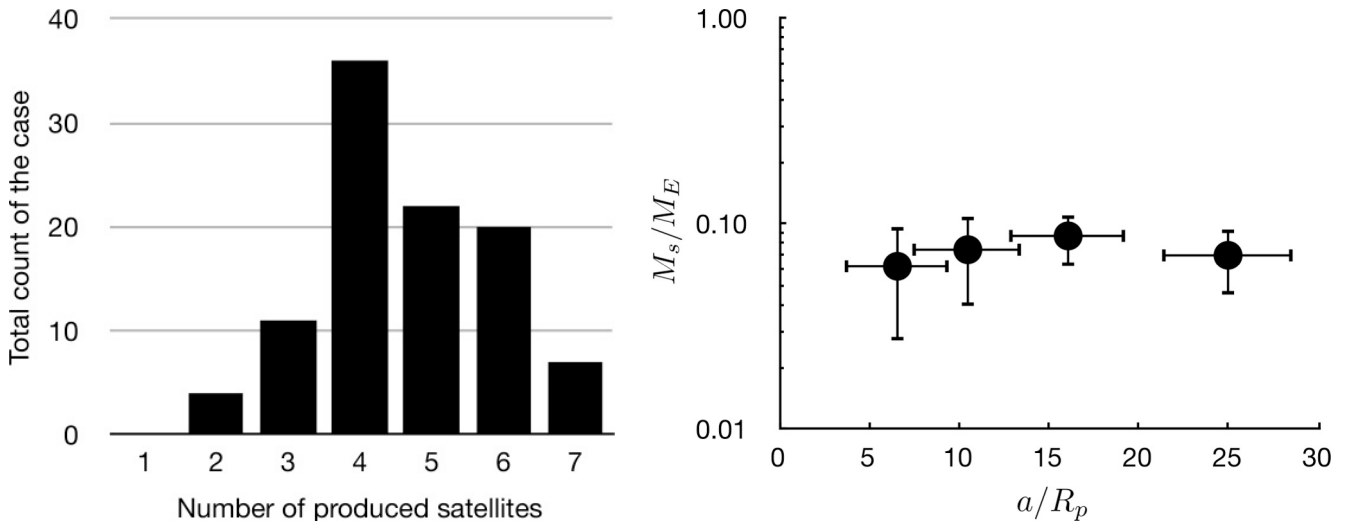

**Figure 2**: Results after 100 simulations of moon formation around a 10 $M_{Jup}$ planet. Left panel: Multiplicity distribution of the produced satellite systems for satellite masses $M_s > 10^{-2}\,M_\oplus$. Right panel: Averaged $M_s$ and semi-major axis ($a$) are shown as filled circles, their standard deviations are indicated by error bars for the 36 four-satellite-systems.

Sasaki et al. (2010) explained how these different final formation phases between Jupiter and Saturn produce the significantly different architectures of their satellite systems, based on the actively supplied gaseous accretion protosatellite disk model (Canup and Ward, 2002). They applied the population synthesis planet formation model (Ida and Lin, 2004, 2008) to simulate growth of protosatellites through accretion of satellitesimals and their inward orbital migration caused by tidal interactions with the gas disk. The evolution of the gas surface density of the protosatellite disk is analytically given by a balance between the infall and the disk accretion. The surface density of satellitesimals is consistently calculated with accretion by protosatellites and supply from solid components in the incident gas. They assumed that the solid component was distributed over many very small dust grains so that the solids can be delivered to the protosatellite disk according to the gas accretion to the central planet. The model includes type I migration and regeneration of protosatellites in the regions out of which preceding runaway bodies have migrated leaving many satellitesimals. Resonant trapping of migrating protosatellites is also taken into account. If the inner disk edge is set, the migration is halted there and the migrated protosatellites are lined up in resonances from the inner edge to the outer regions. When the total mass of the trapped satellites exceeds the disk mass, the halting mechanism is not effective, such that they release the innermost satellite into the planet.

Sasaki et al. (2010) showed that in the case of the Jovian system, a few satellites of similar masses were likely to remain in mean motion resonances. These configurations form by type I migration, temporal stopping of the migration near the disk inner edge, and quick truncation of gas infall by gap opening in the solar nebula. On the other hand, the Saturnian system tended to end up with one dominant body in its outer regions caused by the slower decay of gas infall associated with global depletion of the solar nebula. Beyond that, the compositional variations among the satellites was consistent with observations with more rocky moons close to the planet and more water-rich moons in wider orbits.

### 3.1.3 Formation of massive exomoons in planetary disks

Ultimately, we want to know whether much more massive satellites, for example the size of Mars, can form around extrasolar giant planets. If the Canup and Ward (2006) mass scaling law is universal, these massive satellites could exist around super-Jovian gas giants. To test this hypothesis, we applied the population synthesis satellite formation model of Sasaki et al. (2010) to a range of hypothetical "super-Jupiters" with tenfold the mass of Jupiter. The results from 100 runs are depicted in Fig. 2, showing that in about 80% of all cases, four to six large bodies are formed. In the right panel, we show the averaged semi-major axes (abscissa) and masses (ordinate), along with the 1σ standard deviations, of those 36 systems that contain four large satellites. In these four-satellite-systems, the objects reach roughly the mass of Mars, and they are composed of rocky materials as they form in massive protosatellite disks with high viscous heating. We conclude that massive satellites around extrasolar gas giants can form in the circumplanetary disk and that they can be habitable if they orbit a giant planet in the stellar habitable zone.

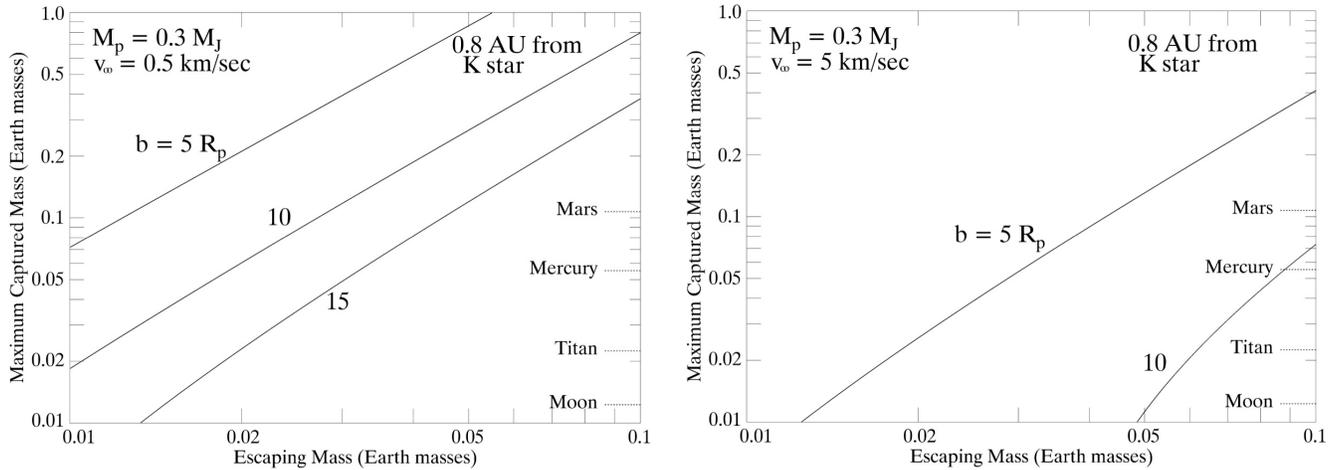

**Figure 3**: Maximum captured mass (ordinate) as a function of escaping mass (abscissa) and encounter distance $b = 5\ R_p$, 10 $R_p$, and 15 $R_p$ (contours). The curves are calculated from Eq. (3) with the planetary mass set to 0.3 $M_{Jup}$ and the distance from the K star $a_\star = 0.8$ AU in both panels. The encounter speed at infinity $v_\infty = 0.5$ km/s in the left panel and $v_\infty = 5$ km/s in the right panel.

### 3.2 Formation by capture

Other ways of forming giant moons include collision and capture, both of which require strong dynamical mixing of giant planets and terrestrial objects in a protoplanetary disk. A gas giant planet can migrate both inward and outward through a protoplanetary disk once it grows large enough to clear a gap in the disk. The migrating gas giant is then able to sweep through terrestrial material, as is thought to have occurred in the Solar System (Gomes et. al, 2005), possibly enabling close encounters that might result in the capture of an over-sized moon. In all capture scenarios, the approaching mass must be decelerated below the planet's escape velocity and inserted into a bound, highly elliptical orbit that can later be circularized by tides (Porter and Grundy, 2011). The deceleration may result from the impactor striking the planet, a circumplanetary disk (McKinnon and Leith, 1995), or a fully formed satellite, but in each of these cases, the intercepting mass – gas or solid – must be comparable to the mass of the impactor to make a significant change in its encounter trajectory. For an impact with the planet to work, the impactor must tunnel deep enough through a planet's outer atmosphere to release enough energy but still shallow enough to avoid disintegration. An impact with a massive satellite already in orbit around the planet might work if the collision is head-on and the impactor-to-satellite mass ratio is smaller than a few, which makes this formation scenario unlikely.

Irrespective of the actual physical reason or history of a possible capture, Porter and Grundy (2011) followed the orbital evolution of captured Earth-mass moons around a range of giant planets orbiting in the HZ of M, G, and F stars. Stellar perturbations on the planet-satellite orbit were treated with a Kozai Cycle and Tidal Friction model, and they assumed their satellites to start the planetary entourage in highly elliptical orbits with eccentricities $e_{ps} > 0.85$ and apoapses beyond 0.8 times the planetary Hill radius

$$R_{\text{Hill,p}} = \left(\frac{M_p}{3M_\star}\right)^{1/3} a_{\star p}$$
(1)

where $M_\star$ is the stellar mass, and $a_{\star p}$ is the orbital semi-major axis of the star-planet system. Initial orbital inclinations and moon spin states were randomized, which allowed investigations of a potential preference for prograde or retrograde orbital stability. Most importantly, their results showed that captured exomoons in stellar HZs tend to be more stable the higher the stellar mass. While about 23% of their captured satellites around Neptune- and Jupiter-sized planets in the HZs of an M0 star remained stable, roughly 45% of such planet-moon binaries in the HZ of a Sun-like star survived and about 65% of similar scenarios in the HZ of an F0 star stabilized over one billion years. This effect is related to the extent of the planet's Hill sphere, which scales inversely proportional to stellar distance while stellar illuminations goes with one over distance squared. No preference for either pro- or retrograde orbits was found. Typical orbital periods of the surviving moons were 0.9, 2.1, and 3.6 days for planet-moon binaries in the HZs of M, G, and F stars, respectively.

A plausible scenario for the origin of these captures was recently discussed by Williams (2013, W13 hereafter), who showed that a massive moon could be captured if it originally belonged to a binary-terrestrial object (BTO) that was tidally disrupted from a close-encounter with a gas-giant planet. During the binary-exchange interaction, one of the BTO members is ejected while the other is captured as a moon. In fact, it is this mechanism that gives the most reasonable formation scenario for Neptune's odd principal moon Triton (Agnor and Hamilton, 2006). The first requirement for a successful capture is for the BTO to actually form in the first place. The second requirement is to pass near enough to the planet (inside five to ten planet radii) to be tidally disrupted. This critical distance is where the planet's gravity exceeds the self-gravity of the binary, and is expressed in W13 as

$$\left(\frac{b}{R_{\mathrm{p}}}\right) \lesssim \left(\frac{a_{\mathrm{B}}}{R_1}\right)\left(\frac{3\rho_{\mathrm{p}}}{\rho_1}\right)^{1/3} \quad, \tag{2}$$

where the subscript "1" refers to the more massive component of the binary, "p" refers to the planet, and $a_{\mathrm{B}}$ is the binary separation. Setting $a_{\mathrm{B}}/R_1 = 10$, and using densities $\rho_{\mathrm{p}}$ and $\rho_1$ appropriate for gaseous and solid planets, respectively, yields $b \approx 9.3 \, R_{\mathrm{p}}$, which shows that the encounter must occur deeper than the orbits of many of the major satellites in the Solar System.

A second requirement for a successful binary-exchange is for the BTO to rotate approximately in the encounter plane as it passes the planet. The binary spin thereby opposes the encounter velocity of the retrograde-moving binary mass so that its final velocity may drop below the escape velocity from the planet if the encounter velocity of the BTO at infinity is small, say ≲ 5 km/s. Finally, the distance and mass of the host star are important because they determine how large of a satellite orbit is stable given the incessant stellar pull. Implementing these physical and dynamical requirements yields an analytic upper limit on the captured mass $m_1$ as a function of the escaping mass $m_2$, in addition to other parameters such as planet mass $M_{\mathrm{p}}$, impact parameter $b$, encounter velocity $v_{\mathrm{enc}}$, and the periapsis velocity $v_{\mathrm{peri}}$ of the newly captured mass (W13). Expressed in compact form, this relation is

$$m_1 < 3M_{\mathrm{p}}\left(\frac{G\,m_2\,\pi}{2\,b\,v_{\mathrm{enc}}(v_{\mathrm{enc}} - v_{\mathrm{peri}})}\right)^{3/2} - m_2 \quad, \tag{3}$$

with $G$ as Newton's gravitational constant and secondary expressions for $v_{\mathrm{enc}}$ and $v_{\mathrm{peri}}$ given in Eqs. (4) and (5) of W13.

It is apparent from this expression that the heaviest moons (large $m_1$) will form from the closest encounters (small $b$) occurring at low velocity (small $v_{\mathrm{enc}}$). The dependence on planet mass is not as straightforward. It appears from Eq. (3) that larger planets should capture heavier masses, but the encounter velocity and periapsis velocity in the denominator both increase with $M_{\mathrm{p}}$, making the dependence on planet mass inverted. This is borne out in Fig. 7 of W13, which shows the size of the captured mass to decrease as planet mass increases with the impact parameter $b$ held constant. Therefore, in general, it is easier to capture a moon around a Saturn- or Neptune-class planet than around a Jupiter or a super-Jupiter because the encounter speeds tend to be smaller.

As detailed in W13, (see his Figs. 3 to 7), moons the size of Mars or even Earth are possibly formed if the ratio of captured mass to escaping mass is not too large. The limiting ratio depends on encounter details such as the size and proximity of both the planet and the star, as well as the encounter velocity. Yet, a ratio > 10:1 could yield a moon as big as Earth around a Jupiter at 2 AU from the Sun (W13, right side of Fig. 5 therein). Mars-sized moons can possibly form by ejecting a companion the size of Mercury or smaller.

Expanding the cases considered in W13, we assume here a Saturn-mass planet in the habitable zone (0.8 AU) around a K-star and vary the encounter distance $b$ as well as the encounter velocity to examine whether velocities as large as 5 km/s make a binary-exchange capture impossible. According to the left panel of Fig. 3, losing an object the size of Mercury (≈ 0.05 $M_{\oplus}$) in a 0.5 km/s encounter would result in the capture of a Mars-sized moon, with tidal disruption occurring as far as 15 $R_{\mathrm{p}}$ from the planet. The right panel of Fig. 3 shows that such an exchange is also possible at ten times this encounter velocity, provided the approach distance is reduced by a factor of three. This is a promising result given the broad range of encounter speeds and approach distances expected between dynamically interacting planets in developing planetary systems. To sum up, close encounters of binary-terrestrial objects can reasonably provide a second formation channel for moons roughly the mass of Mars.

## 4. Orbital Dynamics

Once the supply of the circumplanetary disk with incident gas and dust has ceased several million years after the formation of the planetary system or once a satellite has been captured by a giant planet, orbital evolution will be determined by the tidal interaction between the planet and the moons, gravitational perturbations among multiple satellites, the gravitational pull from the star, and perturbations from other planets. These effects give rise to a range of phenomena, such as spin-orbit resonances, mean motion resonances among multiple satellites, chaos, ejections, and planet-satellite mergers.

As an example, a giant planet orbiting its host star closer than about 0.5 AU will have its rotation frequency $\Omega_p$ braked and ultimately synchronized with its orbital motion $n_{\star p}$ around the star. This rotational evolution, $\Omega_p = \Omega_p(t)$ ($t$ being time), implies an outward migration of the planet's corotation radius, at which $\Omega_p = n_{ps}$, with $n_{ps}$ as the orbital mean motion of a satellite around the planet. Due to the exchange of orbital and rotational momentum via the planet's tidal bulge raised by a moon, satellites inside the corotation radius will spiral towards the planet and perhaps end in a collision (see Phobos' fall to Mars; Efroimsky and Lainey, 2007), while satellites beyond the corotation radius will recede from their planets and eventually be ejected (Sasaki et al., 2012). Hence, the evolution of a planet's corotation radius due to stellar-induced tidal friction in the planet affects the stability of moon systems. Barnes and O'Brien (2002) considered Earth-mass moons subject to an incoming stellar irradiation similar to the solar flux received by Earth and found that these satellites can follow stable orbits around Jupiter-like planets if the host star's mass is greater than 0.15 $M_\odot$, with $M_\odot$ as one solar mass. From another perspective, satellite systems around giant planets that orbit their stars beyond 0.6 AU should still be intact after 5 Gyr. Cassidy et al. (2009), however, claimed that an Earth-sized moon could even follow a stable orbit around a hot Jupiter if the planet is rotationally synchronized to its orbit around the star. Then the tidal forcing frequencies raised by the moon on the planet would be in a weakly dissipative regime, allowing for a slow orbital evolution of the moon. Combining the results of Barnes and O'Brien (2002) with those of Domingos et al. (2006), Weidner and Horne (2010) concluded that 92% of the transiting exoplanets known at that time (almost all of which have masses between that of Saturn and a few times that of Jupiter) could not have prograde satellites akin to Earth's Moon. Further limitations on the orbital stability of exomoons are imposed by the migration-ejection instability, causing giant planets in roughly 1-day orbits to lose their moons during migration within the circumstellar disk (Namouni, 2010).

Investigations on the dynamical stability of exomoon systems have recently been extended to the effects of planet-planet scattering. Gong et al. (2013) found that planetary systems whose architectures are the result of planet-planet scattering and mergers should have lost their initial satellite systems. Destruction of moon systems would be particularly effective for scattered hot Jupiters and giant planets on eccentric orbits. Most intriguingly, in their simulations, the most massive giant planets were not hosts of satellite systems, if these planets were the product of former planet-planet mergers. In a complimentary study, Payne et al. (2013) explored moons in tightly packed giant planet systems, albeit during less destructive planet-planet encounters. For initially tight planet configurations, that is, planet architectures that avoid planet-planet mergers or ejections, they found that giant exoplanets in closely packed systems can very well harbor exomoon systems.

Orbital effects constrain the habitability of moons. Heller (2012) concluded that stellar perturbations would force exomoons into elliptical orbits, thereby generating substantial amounts of tidal heat in the moons. Giant planets in the HZs around low-mass stars have small Hill radii, and so their moons would necessarily have small semi-major axes. As a consequence of both the requirement for a close orbit and the stellar-induced orbital eccentricities, Mars- to Earth-sized exomoons in the HZs of stars with masses below 0.2 to 0.5 $M_\odot$ would inevitably be in a runaway greenhouse state due to their intense tidal heating.

To illustrate the evolution in satellite systems in the following, we simulate the orbital evolution of hypothetical exomoon systems and discuss secular and tidal processes in more detail. Stellar perturbations are neglected. At first, we consider single satellite systems and then address multiple satellite systems.

### 4.1 Orbital evolution in single satellite systems

Imagine a hypothetical Earth-mass satellite orbiting a Jupiter-mass planet. Both the tides raised on the planet by the satellite (the planetary tide) and the tides raised on the satellite by the planet (the satellite tide) can dissipate energy from this two-body system. For our computations, we use the constant-time-lag model, which is a standard equilibrium tidal model (Hut, 1981). As in the work of Bolmont et al. (2011, 2012), we solve the secular equations for this system assuming tidal dissipation in the moon to be similar to that on Earth, where $\Delta\tau_\oplus = 638$s is the time lag between Earth's tidal bulge and the line connecting the two centers of mass (Neron de Surgy and Laskar, 1997), and $k_{2,\oplus} \times \Delta\tau_\oplus = 213$ s, with $k_{2,\oplus}$ as the Earth's second-degree tidal Love number. For Jupiter, we use the value given by Leconte et al. (2010) ($k_{2,Jup} \times \Delta\tau_{Jup} = 2.5 \times 10^{-2}$ s) and include the evolution of the planetary radius $R_{Jup}$, of $k_{2,J}$, and of the radius of gyration squared ($r_{g,Jup}^2 = I_{Jup}/(M_{Jup} R_{Jup}^2)$), where $I_{Jup}$ is Jupiter's moment of inertia, following the evolutionary model of Leconte and Chabrier (2012, 2013). This model was scaled so that the radius of Jupiter at the age 4.5 Gyr is equal to its present value.

The left panel of Fig. 4 shows the evolution of the semi-major axes (top panel) and spin-orbit misalignments, or "obliquities" (bottom panel), of various hypothetical Earth-mass satellites around a Jupiter-mass planet for different initial semi-major axes. The orbital eccentricity $e_{ps}$ is zero and the moon is in synchronous rotation, so once the obliquity is zero only the planetary tide influences the evolution of the system. Not shown in these diagrams is the rotation of the Jupiter, which spins up due to contraction and which is modified by the angular momentum transfer from the satellite's orbit (Bolmont et al., 2011). Figure 4 shows that the corotation radius (ensemble of dashed blueish lines) – defined as the distance at which the satellite's orbital frequency equals the planetary rotation frequency – shrinks due to the spin-up of the planet. A satellite initially interior to the corotation radius migrates inward, because the bulge of the Jupiter is lagging behind the position of the satellite, and eventually falls onto the planet. A satellite initially exterior to the corotation radius migrates outward. Most notably, an Earth-mass satellite can undergo substantial tidal evolution even over timescales that span the age of present Jupiter. The bottom left panel of Fig. 4 shows the evolution of the obliquities of the satellites, whose initial value is set to 11.5° for all tracks. In all cases, the obliquities rapidly eroded to zero, in less than a year for a very close-in satellite and in a few $10^4$ yr for a satellite as far as $7 \times 10^{-3}$ AU. Thus, in single satellite systems, satellite obliquities are likely to be zero. The timescale of evolution of the satellite's rotation period is similar to the obliquity evolution timescale, so a satellite gets synchronized very quickly (for $e_{ps} \approx 0$) or pseudo-synchronized (if $e_{ps}$ is substantially non-zero).

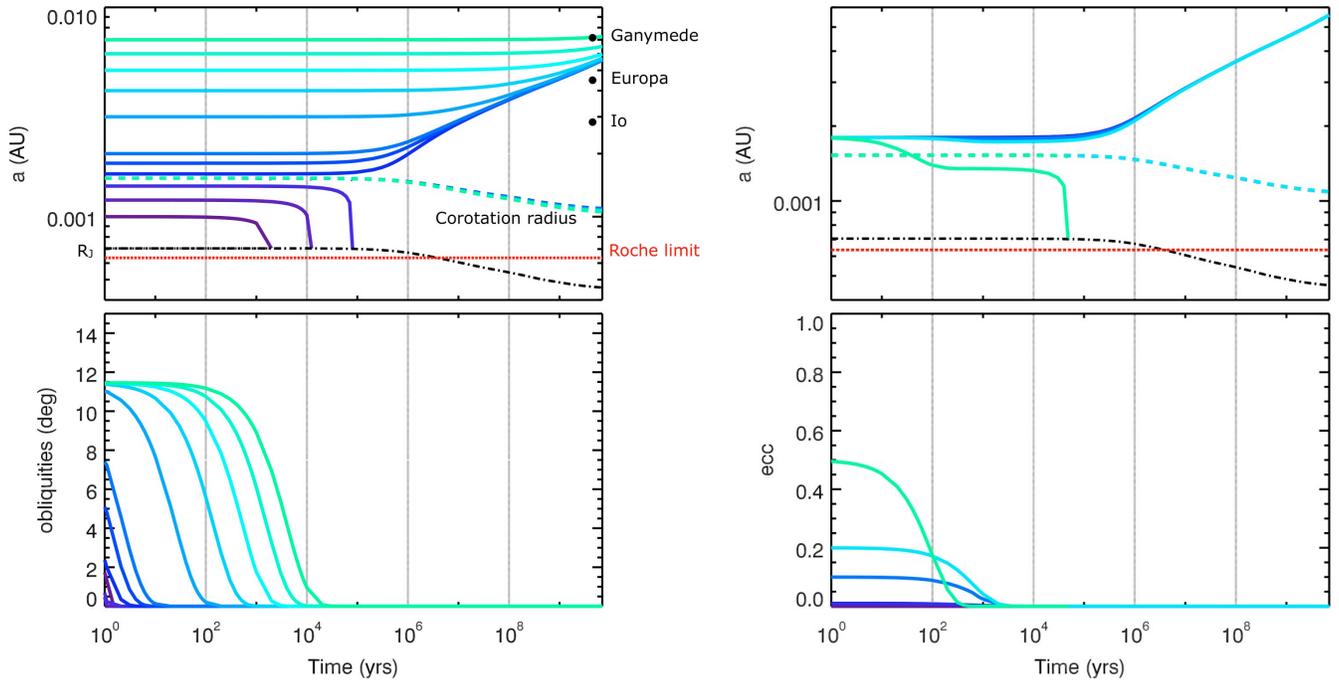

**Figure 4**: Tidal evolution of an Earth-sized moon orbiting a Jupiter-sized planet. Left panels: Semi-major axis (top) and obliquity (bottom) evolution for different initial semi-major axes, while all other initial parameters are equal. The black dashed dotted line in the top panel represents the planetary radius; four overlapping dashed lines indicate the corotation radii. A red dashed line represents the Roche limit. Right panels: Semi-major axis (top) and eccentricity (bottom) evolution for the same system but for different initial eccentricities.

The right panel of Fig. 4 shows the evolution of semi-major axes (top panel) and eccentricities (bottom panel) of the same Earth-Jupiter binaries as in the left panel, but now for different initial eccentricities. In the beginning of the evolution, the eccentricity is damped by both the planetary and the satellite tides in all simulations. What is more, all solid tracks start beyond their respective dashed corotation radius, so the satellites should move outward. However, the satellite corresponding to the light blue curve undergoes slight inward migration during the first $10^5$ years, which is due to the satellite tide that causes eccentricity damping. There is a competition between both the planetary and the satellite tides, and when the eccentricity is > 0.08 then the satellite tide determines the evolution causing inward migration. But if $e_{ps}$ < 0.08, then the planetary tide dominates and drives outward migration (Bolmont et al., 2011). The planet with initial $e_{ps}$ = 0.5 undergoes rapid inward migration such that it is interior to the corotation radius when its eccentricity approaches zero. Hence, it then transfers orbital angular momentum to the planet until it merges with its host.

### 4.2 Orbital evolution in multiple-satellite systems

We consider now a system of two satellites orbiting the same planet. One moon is a Mars-mass satellite in a close-in orbit at 5 $R_{Jup}$, the other one is an Earth-mass satellite orbiting farther out, between 10 and 50 $R_{Jup}$. All simulations include the structural evolution of the Jupiter-mass planet as in the previous subsection. We use a newly developed code, based on the *Mercury* code (Chambers, 1999), which takes into account tidal forces between both satellites and the planet as well as secular interaction. We also consider tides raised by each of the satellites on the planet, assuming these planetary bulges are independent, and the tide raised by the planet in each of the two satellites. The code consistently computes the orbital evolution as well as the evolution of the rotation period of the three bodies and their obliquities (Bolmont et al., 2014). The inner satellite is assumed to have an initial eccentricity of 0.05, an orbital inclination of 2° with respect to the planet's equatorial plane, and an obliquity of 11.5°. The outer satellite has an initial eccentricity of 0.1, an inclination of 5° to the planet's equator, and an obliquity of 7°. Tidal dissipation parameters for the satellite and the planet are assumed as above.

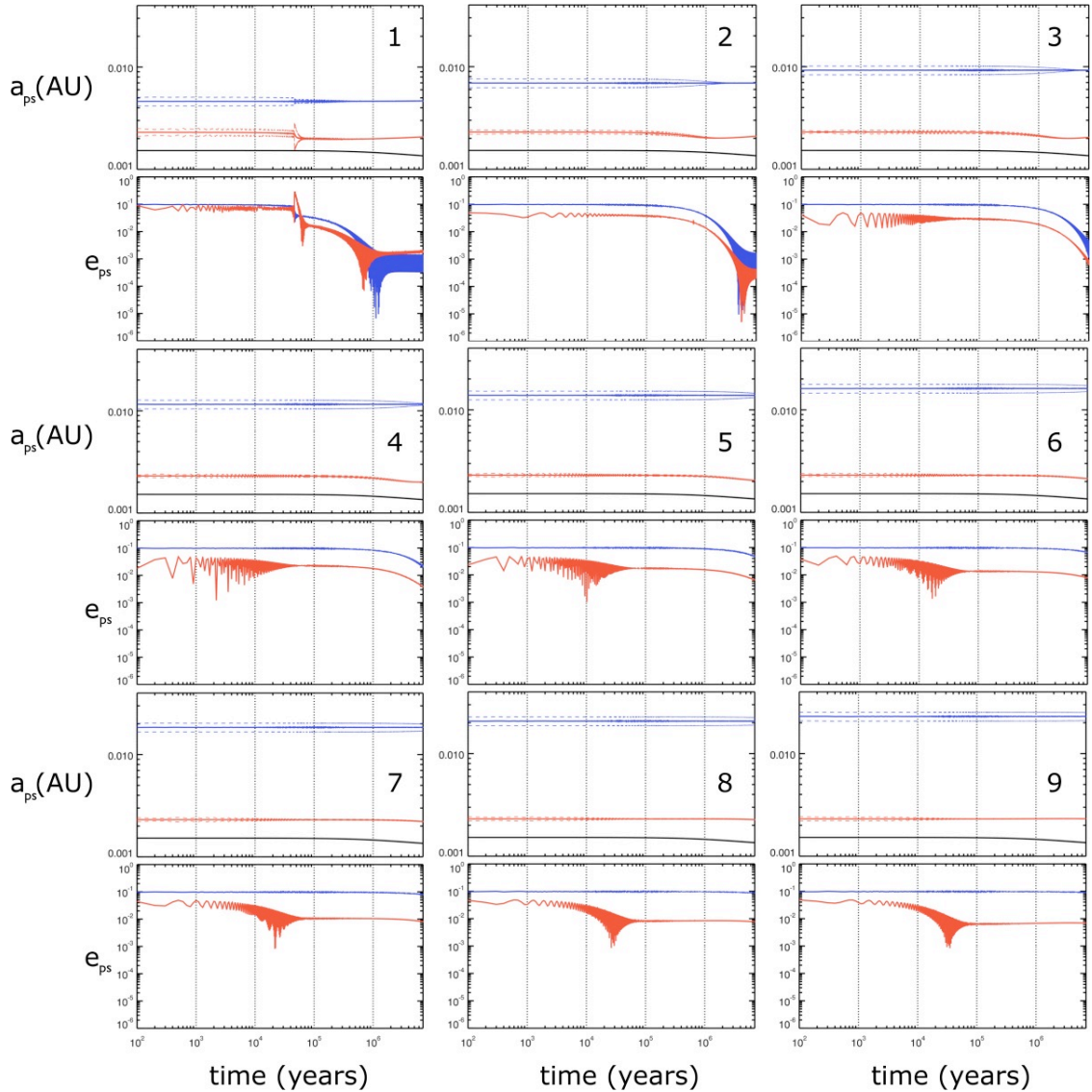

**Figure 5**: Evolution of the semi-major axes ($a_{ps}$) and eccentricities ($e_{ps}$) of the two satellites orbiting a Jupiter-mass planet. Red lines correspond to the inner Mars-mass satellite, blue lines to the outer Earth-mass satellite, and black lines represent the corotation distance. Each panel depicts a different initial distance for the outer satellite, increasing from panels 1 to 9.

Figure 5 shows the results of these simulations, where each panel depicts a different initial semi-major axis of the outer moon. In all runs, the outer massive satellite excites the eccentricity of the inner moon. Excitations result in higher eccentricities if the outer satellite is closer in. Both eccentricities decrease at a similar pace, which is dictated by the distance of the outer and more massive satellite: the farther away the outer satellite, the more slowly the circularization. At the same time, the inner smaller satellite undergoes inward migration due to the satellite tide. When its eccentricity reaches values below a few $10^{-3}$, the planetary tide causes outward migration because the moon is beyond the corotation radius (panels 1 to 3). In panels 4 to 9, $e_{ps}$ of the outer moon decreases on timescales > 10 Myr, so the system does not reach a state in which the planetary tide determines the evolution to push the inner satellite outward.

Panel 1 of Fig. 5 shows an instability between $10^4$ and $10^5$ yr. The inner satellite eccentricity is excited to values close to the outer satellite eccentricity. At a few $10^4$ yr, $e_{ps}$ of the inner satellite is excited to even higher values, but the satellite tide quickly damps it. This rapid circularization is accompanied by a substantial decrease of the inner moon's semi-major axis. Once the eccentricity reaches values of $10^{-2}$, evolution becomes smoother. Both eccentricities then decrease and dance around a kind of "equilibrium" at $e_{ps} \approx 10^{-3}$. On even longer timescales, both eccentricities are expected to decrease further towards zero (Mardling, 2007).

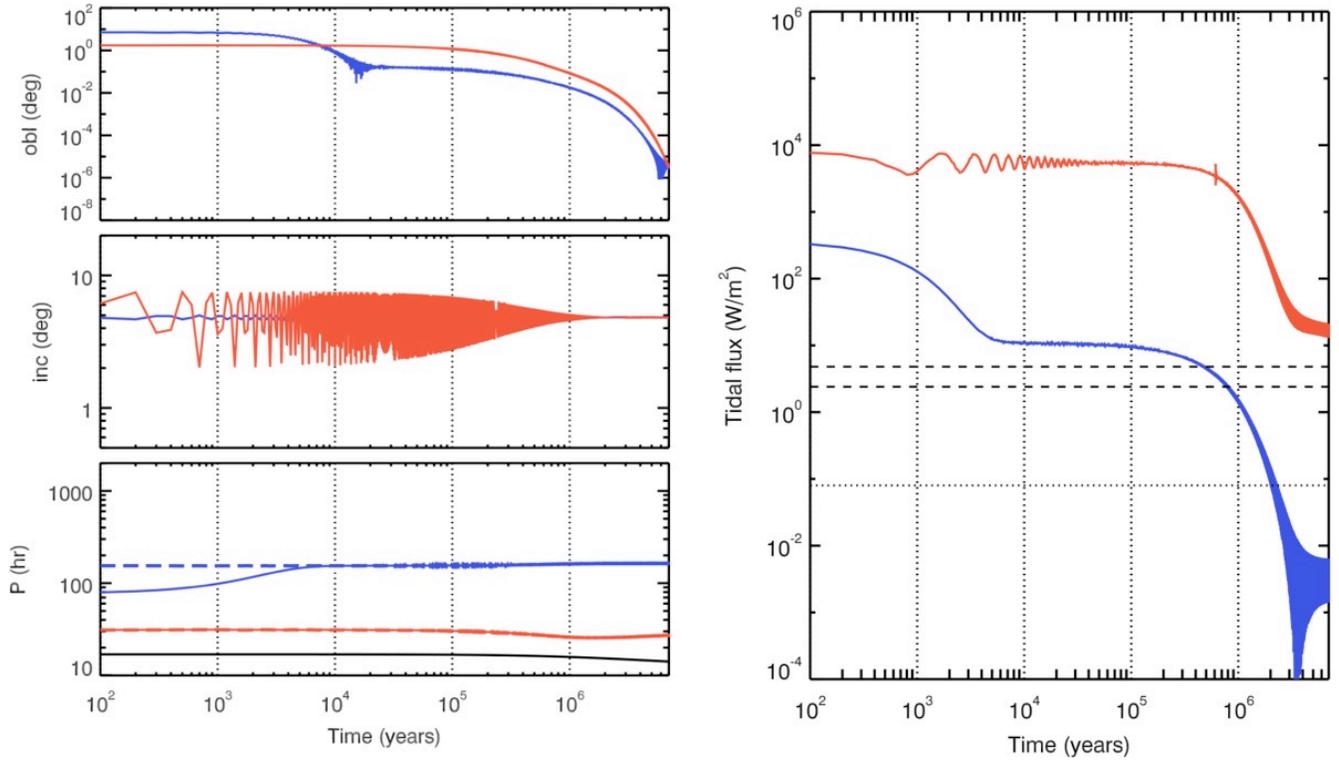

**Figure 6**: Evolution of the obliquity (top left), inclination (center left), rotation period (bottom left), and internal heat flux (right) of a Mars- (red lines) and an Earth-like (blue lines) satellite orbiting a Jupiter-mass planet (configuration of panel 2 in Fig. 9). In the bottom left panel, the black line represents the rotation period of the planet, and the dashed lines correspond to the pseudo-synchronization period of the satellites. In the right panel, the dotted black horizontal line corresponds to the internal heat flux of Earth (0.08 W/m$^2$), the dashed black lines correspond to tidal surface heating on Io (2.4 - 4.8 W/m$^2$).

Figure 6 shows the evolution of the obliquities, inclinations, rotation periods, and tidal surface heat flux for both satellites in the scenario of panel 2 of Fig. 5. The inner satellite reaches pseudo-synchronization in < 100 yr, during that time its obliquity decreases from initially 11.5° to about 2°. The outer satellite reaches pseudo-synchronization in a few 1000 yr, while its obliquity decreases to a few $10^{-1}$ degrees within some $10^4$ yr. Once pseudo-synchronization is reached on both satellites, their obliquities slowly decrease to very low values between $10^4$ and $10^7$ yr in Fig. 6. The inclination of the small inner satellite oscillates around the value of the more massive outer one, whose inclination does not vary significantly during the first $10^7$ yr because the planetary tides are not efficient enough to adjust the orbital planes to the planet's equatorial plane.

Moving on to the tidal heat flux in the right panel of Fig. 6, note that the internal heat flux through the Earth's surface is around 0.08 W/m$^2$ (Pollack et al., 1993) and Io's value is between 2.4 and 4.8 W/m$^2$ (Spencer et al., 2000). Initially, the tidal heat flux is extremely large in both satellites. While being pseudo-synchronized during the first $\approx 10^4$ yr, the tidal heat flux of the outer satellite decreases from some $10^2$ to about 10 W/m$^2$. After $10^4$ yr, the decay of tidal heating in both satellites is due to the circularization and tilt erosion (Heller et al., 2011b) on a 10 Myr timescale. The tidal surface heat flux of the outer satellite is about an order of magnitude lower than Earth's internal heat flux through its surface, but the tidal heat going through a surface unit on the inner moon is several times the tidal surface heat flux on Io, probably causing our Mars-mass test moon to show strong tectonic activity for as long as its first 100 Myr after formation.

The long-term evolution of this system might lead to interesting configurations. For example, the outward migration of the inner satellite decreases the distance between the two satellites, eventually resulting in an orbital resonance (Bolmont et al., 2014). In such a resonance, the eccentricities of both satellites would be excited and thereby reignite substantial tidal heating. Not only would such a resurgence affect the satellites' potentials to be, remain, or become habitable, but their thermal emission could even prevail against the host planet's thermal emission and make the moon system detectable by near-future technology (Section 6.4).

## 5. Effects of Orbital and Planetary Evolution on Exomoon Habitability

Predictions of exomoons the size of Mars orbiting giant planets (Section 3) and the possible detection of exomoons roughly that size with the *Kepler* space telescope or near-future devices (Section 6) naturally make us wonder about the habitability of these worlds. Tachinami et al. (2011) argued that a terrestrial world needs a mass ≳ 0.1 $M_\oplus$ to sustain a magnetic shield on a billion-year timescale, which is necessary to protect life on the surface from high-energy stellar and interstellar radiation. Further constraints come from the necessity to hold a substantial, long-lived atmosphere, which requires satellite masses $M_s$ ≳ 0.12 $M_\oplus$ (Williams et al., 1997; Kaltenegger, 2000). Tectonic activity over billions of years, which is mandatory to entertain plate tectonics and to promote the carbon-silicate cycle, requires $M_s$ ≳ 0.23 $M_\oplus$ (Williams et al., 1997). Hence, formation theory, current technology, and constraints from terrestrial world habitability point towards a preferred mass regime for habitable moons that can be detected with current or near-future technology, which is between 0.1 and 0.5 $M_\oplus$.

### 5.1 The global energy flux on moons

While these mass limits for a habitable world were all derived from assuming a cooling, terrestrial body such as Mars, exomoons have an alternative internal heat source that can retard their cooling and thus maintain the afore-mentioned processes over longer epochs than in planets. This energy source is tidal heating, and its effects on the habitability of exomoons have been addressed several times in the literature (Reynolds et al, 1987; Scharf, 2006; Henning et al, 2009; Heller, 2012; Heller and Barnes, 2013, 2014; Heller and Zuluaga, 2013; Heller and Armstrong, 2014). In their pioneering study, Reynolds et al. (1987) point out the remarkable possibility that water-rich extrasolar moons beyond the stellar HZ could be maintained habitable mainly due to tidal heating rather than stellar illumination. Their claim was supported by findings of plankton in Antarctica lakes, which require an amount of solar illumination that corresponds to the flux received at the orbit of Neptune. Heller and Armstrong (2014) advocated the idea that different tidal heating rates allow exomoons to be habitable in different circumplanetary orbits, depending on the actual distance of the planet-moon system from their common host star.

Assuming an exomoon were discovered around a giant planet in the stellar HZ, then a first step towards assessing its habitability would be to estimate its global energy flux $\bar{F}_s^{\mathrm{glob}}$. Were the moon to orbit its planet too closely, then it might be undergoing a moist or runaway greenhouse effect (Kasting, 1988; Kasting et al., 1993) and be temporarily uninhabitable or even desiccated and uninhabitable forever (Heller and Barnes, 2013, 2014). In addition to the orbit-averaged flux absorbed from the star ($\bar{f}_\star$), the moon absorbs the star's reflected light from the planet ($\bar{f}_r$), the planet's thermal energy flux ($\bar{f}_t$), and it can undergo substantial tidal heating ($h_s$, the amount of tidal heating going through a unit area on the satellite surface). The total global average flux at the top of a moon's surface is then given by

$$
\begin{aligned}
\bar{F}_s^{\mathrm{glob}} &= \bar{f}_\star \;+\; \bar{f}_r \;+\; \bar{f}_t \;+\; h_s \;+\; W_s \\
&= \frac{L_\star\,(1-\alpha_{s,\mathrm{opt}})}{4\pi a_{\star p}^2 \sqrt{1-e_{\star p}^2}}\left(\frac{x_s}{4}+\frac{\pi R_p^2 \alpha_p}{f_s\,2\,a_{ps}^2}\right) \;+\; \frac{L_p(1-\alpha_{s,\mathrm{IR}})}{4\pi a_{ps}^2 f_s \sqrt{1-e_{ps}^2}} \;+\; h_s \;+\; W_s \\
&= \frac{R_\star \sigma_{\mathrm{SB}}^2 T_{\mathrm{eff},\star}^4\,(1-\alpha_{s,\mathrm{opt}})}{a_{\star p}^2 \sqrt{1-e_{\star p}^2}}\left(\frac{x_s}{4}+\frac{\pi R_p^2 \alpha_p}{f_s\,2\,a_{ps}^2}\right) \;+\; \frac{R_p^2 \sigma_{\mathrm{SB}} T_{p,\mathrm{eff}}^4(1-\alpha_{s,\mathrm{IR}})}{a_{ps}^2 f_s \sqrt{1-e_{ps}^2}} \;+\; h_s \;+\; W_s
\end{aligned}
$$
,
(4)

where $L_\star$ and $L_p$ are the stellar and planetary luminosities, respectively, $\alpha_{s,\mathrm{opt}}$ and $\alpha_{s,\mathrm{IR}}$ the satellite's optical and infrared albedo, respectively (Heller and Barnes, 2014), $\alpha_p$ is the planetary bond albedo, $a_{\star,p}$ and $a_{p,s}$ are the star-planet[19] and planet-satellite semi-major axes, respectively, $e_{\star,p}$ and $e_{p,s}$ the star-planet and planet-satellite orbital eccentricities, respectively, $R_p$ is the planetary radius, $x_s$ the fraction of the satellite orbit that is *not* spent in the planetary shadow (Heller, 2012), $\sigma_{\mathrm{SB}}$ the Stefan-Boltzmann constant, $T_{p,\mathrm{eff}}$ the planet's effective temperature, $h_s$ is the tidal surface heating of the satellite, and $W_s$ denotes contributions from other global heat sources such as primordial thermal energy (or "sensible heat"), radioactive decay, and latent heat from solidification. The planet's effective temperature $T_{p,\mathrm{eff}}$ is a function of long timescales as young, hot giant planets cool over millions of years; and it depends on the planetary surface temperature $T_{p,\mathrm{eq}}$, triggered by thermal equilibrium between absorbed and re-emitted stellar light, as well as on the internal heating contributing a surface component $T_{p,\mathrm{int}}$. Hence, $T_{p,\mathrm{eff}} = ([T_{p,\mathrm{eq}}]^4 + [T_{p,\mathrm{int}}]^2)^{1/4}$ (Heller and Zuluaga, 2013). The factor $f_s$ in the denominator of the terms describing the reflected and thermal irradiation from the planet accounts for the efficiency of the flux distribution over the satellite surface. If the planet rotates freely with respect to the planet, then $f = 4$, and if the satellite is tidally locked with one hemisphere pointing at the planet permanently, then $f = 2$ (Selsis et al, 2007). Another factor of 2 in the denominator of the term related to the stellar light reflected from the planet onto the satellite accounts for the fact that only half of the planet is actually starlit. The split-up of absorbed stellar and planetary illumination in Eq. (4), moderated by the different optical and infrared albedos, has been suggested by Heller and Barnes (2014) and is owed to the different spectral regimes, in which the energy is emitted from the sources and absorbed by the moon.

---

[19] It is assumed that the planet is much more massive than its moons and that the barycenter of the planet and its satellite system is close to the planetary center.

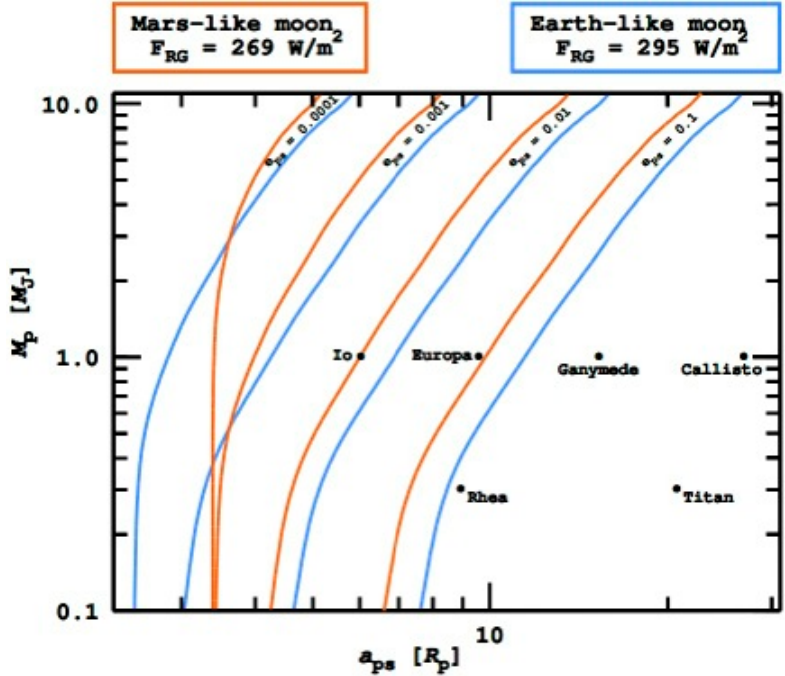

**Figure 7**: Circumplanetary habitable edges (HEs) for a Mars-mass (orange lines) and an Earth-mass (blue lines) exomoon orbiting a range of host planets (masses indicated along the ordinate). All system are assumed to be at 1 AU from a Sun-like star. HEs are indicated for four different orbital eccentricities: $e_{ps} \in \{0.0001, 0.001, 0.01, 0.1\}$. The larger $e_{ps}$, the further away the moons need to be around a given planet to avoid transition into a runaway greenhouse due to extensive tidal heating. Examples for the orbital distances found in the Jovian and Saturnian satellite systems are indicated with labeled dots.

A progenitor version of this model has been applied to examine the maximum variations in illumination that moons in wide circumplanetary orbits can undergo, indicating that these fluctuations can be up to several tens of W/m² (Hinkel and Kane, 2013). Taking into account stellar illumination and tidal heating, Forgan and Kipping (2013) applied a one-dimensional atmosphere model to assess the redistribution of heat in the Earth-like atmospheres of Earth-like exomoons. In accordance with the results of Hinkel and Kane, they found that global mean temperatures would vary by a few Kelvin at most during a circumplanetary orbit. While global flux or temperature variations might be small on moons with substantial atmospheres, the local distribution of stellar and planetary light on a moon can vary dramatically due to eclipses and the moon's tidal locking (Heller, 2012; Heller and Barnes, 2013; Forgan and Yotov, 2014).

Equation (4) can be used to assess an exomoon's potential to maintain liquid reservoirs of water, if $\bar{F}_s^{\mathrm{glob}}$ is compared to the critical flux $F_{\mathrm{RG}}$ for a water-rich world with an Earth-like atmosphere to enter a runaway greenhouse state. We use Eq. (8) of Fortney et al. (2007b) and an Earth-like rock-to-mass fraction of 68% to calculate the two satellites' radii $R_s = 0.5 \ R_\oplus$ and 1 $R_\oplus$, respectively, where $R_\oplus$ symbolizes the radius of Earth. Then, using the semi-analytic atmosphere model of Pierrehumbert (2010, see also Eq. 1 in Heller and Barnes, 2013), we compute $F_{\mathrm{RG}} = 269$ W/m² and 295 W/m² for a Mars-mass and and Earth-mass exomoon, respectively. The closer an exomoon is to its host planet, the stronger tidal heating and the stronger illumination from the planet. Ultimately, there exists a minimum circumplanetary orbit at which the moon transitions into a runaway greenhouse state, thereby becoming uninhabitable. This critical orbit has been termed the circumplanetary "habitable edge", or HE for short (Heller and Barnes, 2013). Giant planets in the stellar HZ will not have an outer habitable edge for their moons, because moons in wide orbits will essentially behave like free, terrestrial planets. Beyond about 15 $R_{\mathrm{Jup}}$, tidal heating will be weak or negligible; illumination from the planet will be very low; eclipses will be infrequent and short compared to the moon's circumplanetary orbital period; and magnetic effects from the planet will be weak. Thus, by definition of the stellar HZ, exomoons could be habitable even at the planetary Hill radius.

As an application of Eq. (4), we assume a planet-moon binary at 1 AU from a Sun-like star. Following Heller and Barnes

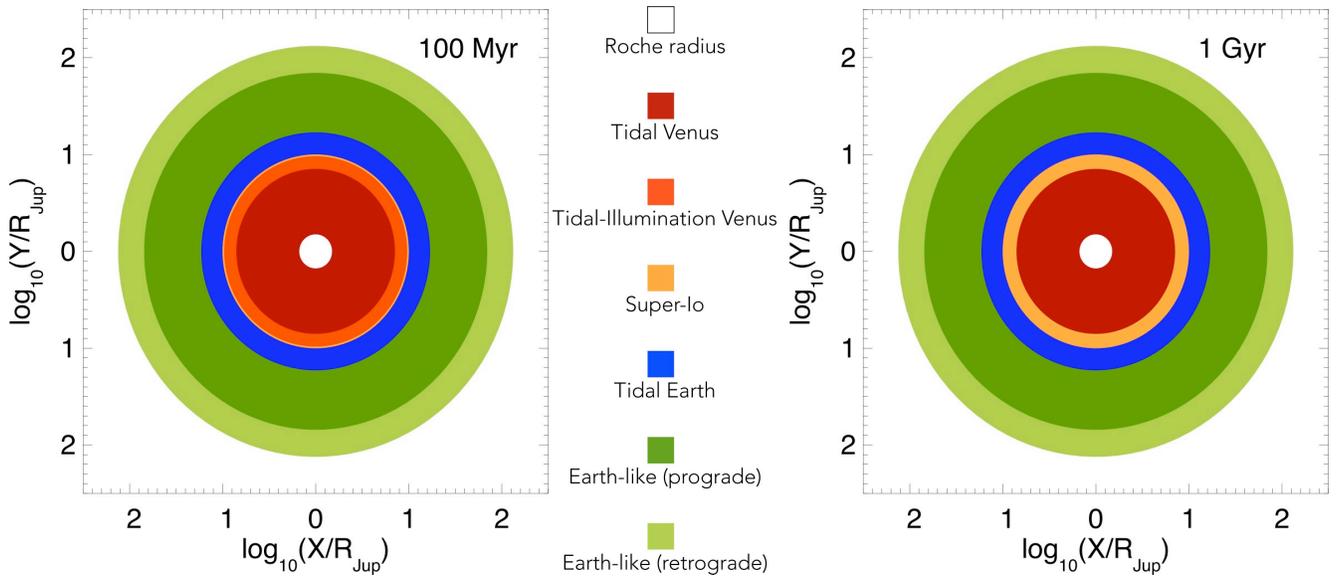

**Figure 8**: Circumplanetary exomoon menageries for Mars-sized satellites around a 10 $M_{Jup}$ host planet at ages of 100 Myr (left panel) and 1 Gyr (right panel). The planet is assumed to orbit in the middle of the HZ of a 0.7 $M_\odot$ star, and the moon orbits the planet with an eccentricity of $10^{-3}$. In each panel, the planet's position is at (0,0), and distances are shown on logarithmic scales. Note that exomoons in a Tidal Venus or a Tidal-Illumination Venus state are in a runaway greenhouse state and thereby uninhabitable.

(2014), we assume that the exomoon's albedo to starlight is $\alpha_{s,opt} = 0.3$, while that due to the planet's thermal emission is $\alpha_{s,IR} = 0.05$. This difference is due to the expectation that the planet will emit primarily in the infrared, where both Rayleigh scattering is less effective and molecular absorption bands are more prominent, decreasing the albedo (Kopparapu et al., 2013). Figure 7 shows the HEs for the Mars- and Earth-like satellites described above (orange and blue lines, respectively), assuming four different orbital eccentricities, $e_{ps} \in \{0.0001, 0.001, 0.01, 0.1\}$ from left to right. Along the abscissa, semi-major axis $a_{ps}$ is given in units of planetary radii $R_p$, which depends on planetary mass $M_p$, which is depicted along the ordinate. To intertwine abscissa and ordinate with one another, we fit a polynomial to values predicted for giant planets with a core mass of 10 $M_\oplus$ at 1 AU from a Sun-like star (Table 4 in Fortney et al., 2007a). Examples for Solar System moons in this mass-distance diagram are indicated with black dots.

For $e_{ps} \gtrsim 0.01$, we find the HEs of the Mars-like satellite are up to a few planetary radii closer to the planet than the HEs for the Earth-like moon. This is due to the strong dependence of tidal heating on the satellite radius, $h_s \propto R_s^3$. Thus, although the Earth-like moon is less resistant against a runaway greenhouse state than the Mars-like satellite, in the sense that 295 W/m² > 269 W/m², it is more susceptible to enhanced tidal heating in a given orbit. We also see a trend of both suites of HEs moving outward from the planet as $M_p$ increases. This is because, firstly, $h_s \propto M_p^3$ for $M_p \gg M_s$ and, secondly, we measure the abscissa in units of planetary radii and radii of super-Jovian planets remain approximately constant around $R_{Jup}$. For $e_{ps} \lesssim 0.01$, we find that the relative positions of the HEs corresponding to the Mars- and Earth-like satellite strongly depend on $M_p$. Again, this is owed to the strong dependence of $h_s$ on $M_p$ and explains why the more resistive Earth-like moon, in terms of a transition to a runaway greenhouse state, can be closer to less massive host planets than the Mars-like moon: in low-eccentricity orbits around low-mass planets, tidal heating is negligible and the HEs depend mostly on additional illumination from the planet.

Orbital eccentricities of exomoons will typically be small due to tidal circularization of their orbits (Porter and Grundy, 2011; Heller and Barnes, 2013). And given that the highest eccentricity among the most massive satellites in the Solar System is 0.0288 for Titan, Fig. 7 suggests that exomoons the mass of Mars or Earth orbiting a giant planet at about 1 AU around a Sun-like star can be habitable if their orbital semi-major axis around the planet-moon barycenter $\gtrsim 10\ R_p$, with closer orbits still being habitable for lower eccentricities and lower-mass host planets.

### 5.2 Runaway greenhouse due to "inplanation"

In this section, we examine the evolution of incident planetary radiation ("inplanation") on the habitability of exomoons, encapsulated in $L_p$ in Eq. (4). Following the prescriptions of Heller and Barnes (2014), we approximate Eq. (4) by considering stellar light, the evolution of inplanation due to the radial shrinking of young gaseous giant planets (Baraffe et al., 2003;

Leconte and Chabrier, 2013), and assuming constant tidal heating. While Heller and Barnes (2014) considered several configurations of star-planet-moon systems, we limit our example to that of a Mars-sized moon (see Section 5.1) around a 10 $M_{Jup}$ planet that orbits a K dwarf in the middle of the HZ. Further, we arbitrarily consider a moon with $e_{ps} = 10^{-3}$, a small value that is similar to those of the Galilean satellites. Heller and Barnes (2014) found that Earth-sized moons orbiting planets less massive than 5 $M_{Jup}$ are safe from desiccation by inplanation, so our example exomoon may be at risk of losing its water from inplanation.

As in the previous subsection, our test exomoon may be considered habitable if the total flux is < 269 W/m². What is more, the duration of the runaway greenhouse is important. Barnes et al. (2013) argued that ≈ 100 Myr are sufficient to remove the surface water of a terrestrial body, based on the pioneering work of Watson et al. (1981). While it is possible to recover habitability after a runaway greenhouse, we are aware of no research that has examined how a planet evolves as energy sources drop below the limit. However, we conjecture that the flux must drop well below the runaway greenhouse limit to sufficiently weaken a $CO_2$ greenhouse to permit stable surface water.

In Fig. 8, we classify exomoons with different surface fluxes as a function of semi-major axis of the planet-moon orbit. We assume satellites of Earth-like properties in terms of overall tidal response, and that 90% of the energy is dissipated in a putative ocean, with the remainder in the solid interior. For this example, we use the "constant-phase-lag" tidal model, as described by Heller et al. (2011a), and assume a tidal quality factor $Q_s$ of 10, along with $k_{2,s} = 0.3$. We classify planets according to the scheme presented by Barnes and Heller (2013) and Heller and Barnes (2014), which the authors refer to as the "exomoon menagerie". If tidal heating alone is strong enough to reach the runaway greenhouse limit, the moon is a "Tidal Venus" (Barnes et al., 2013), and the orbit is colored red. If the tidal heating flux is less, but the total flux is still sufficient to trigger the runaway greenhouse, the moon is a "Tidal-Illumination Venus," and the orbit is orange. These two types of moons are uninhabitable. If the surface flux from the rocky interior is between the runaway limit and that observed on Io (≈ 2 W/m²; Veeder et al., 1994; Spencer et al., 2000), then we label the moon a "Super-Io" (see Jackson et al., 2008; Barnes et al., 2009a, 2009b), and the orbit is yellow. If the tidal heating of the interior is less than Io's but larger than a theoretical limit for tectonic activity on Mars (0.04 W/m²; Williams et al., 1997), then we label the moon a "Tidal Earth," and the orbit is blue. For lower tidal heat fluxes, the moon is considered Earth-like, and the orbit is green. The outer rim of this sphere containing Earth-like moons is determined by stability criteria found by Domingos et al. (2006). Satellites orbiting their planet in the same direction as the planet orbits the star, that is, in a prograde sense, are bound to the planet as far as about 0.49 $R_{Hill,p}$. Moons orbiting their planet in the opposite direction, in a retrograde sense, can follow stable orbits out to 0.93 $R_{Hill,p}$, depending on eccentricities.

The left panel of Fig. 8 shows these classes for a 100 Myr old planet, the right after 1 Gyr. As expected, all boundaries are the same except for the Tidal-Illumination Venus limit, which has moved considerably inward (note the logarithmic scale!). This shrinkage is due to the decrease in inplanation, as all other heat sources are constant[20]. In this case, moons at distances ≤ 10 $R_{Jup}$ are in a runaway greenhouse for at least 100 Myr, assuming they formed contemporaneously with the planet, and thus are perilously close to permanent desiccation. Larger values of planet-moon eccentricity, moon mass, and/or moon radius increase the threat of sterilization. Thus, should our prototype moon be discovered with $a_{ps}$ ≤ 10 $R_{Jup}$, then it may be uninhabitable due to an initial epoch of high inplanation.

### 5.3 Magnetic environments of exomoons

Beyond irradiation and tidal heating, the magnetic environments of moons determine their surface conditions. It is now widely recognized that magnetic fields play a role in the habitability of exoplanetary environments (Lammer et al., 2010). Strong intrinsic magnetic fields can serve as an effective shield against harmful effects of cosmic rays and stellar high energy particles (Grießmeier et al., 2005). Most importantly, an intrinsic magnetic shield can help to protect the atmosphere of a terrestrial planet or moon against non-thermal atmospheric mass loss, which can obliterate or desiccate a planetary atmosphere, see Venus and Mars (Zuluaga et al., 2013). To evaluate an exomoon's habitability, assessing their magnetic and plasmatic environments is thus crucial (Heller and Zuluaga, 2013).

To determine the interaction of a moon with the magnetic and plasmatic environment of its host planet, we need to estimate the size and shape of the planetary magnetosphere. Magnetospheres are cavities within the stellar wind, created by the intrinsic magnetic field of the planet (Fig. 9). The scale of a magnetosphere is given by the distance between the planetary center and the planetary magnetopause. This so-called "standoff radius" $R_S$ has been measured for giant planets in the Solar System (see Table 1). For extrasolar planets, however, $R_S$ can only be estimated from theoretical and semi-empirical models, which predict that $R_S$ scales with the dynamic pressure of the stellar wind $P_{sw} \propto n_{sw} v_{sw}^2$, with $n_{sw}$ as the particle density and $v_{sw}$ as the particle velocity, and with the planetary dipole moment $\mathcal{M}$ as

$$R_S \; \propto \; \mathcal{M}^{1/3} \; \times \; P_{sw}^{-1/\alpha_{mag}} \qquad . \tag{5}$$

---

[20] Without external perturbations, tidal heating should also dissipate as the eccentricity is damped, but here we do not consider orbital evolution.

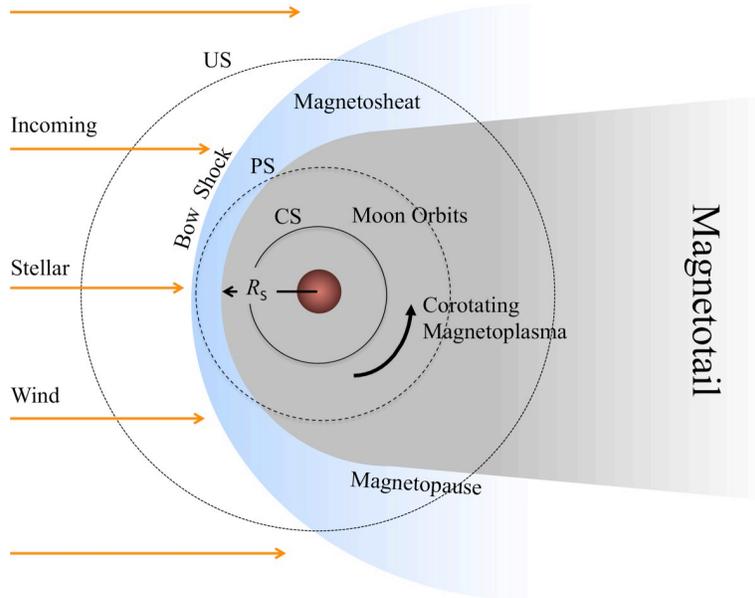

**Figure 9**: Schematic representation of a planetary magnetosphere. Its roughly spherical dayside region has a standoff radius $R_S$ to the planetary center. Between the bow shock and the magnetopause lies the magnetosheat, a region of shocked stellar wind plasma and piled up interplanetary magnetic fields. Inside the magnetosphere, plasma is dragged by the co-rotating magnetic field, thereby creating a particle stream called the "co-rotating magnetoplasma". Moons at orbital distances $< R_S$ (completely shielded, CS) can be subject to this magnetospheric wind and, hence, experience similar effects as unshielded (US) moons exposed to stellar wind. Partially shielded (PS) moons that spend most of their orbital path inside the magnetosphere will experience both effects of interplanetary magnetic fields and stellar wind periodically.

Here, $\alpha_{mag}$ depends on the magnetospheric compressibility, that is, the contribution to internal stress provided by magnetospheric plasma. Values for $\alpha_{mag}$ range from 6 in the case of relatively empty magnetospheres such as that of Earth, Uranus, and Neptune (Zuluaga et al., 2013; Heller and Zuluaga, 2013), to 4 if important magnetospheric plasma sources are present as is the case for Jupiter and Saturn (Huddleston et al., 1998; Arridge et al., 2006). Predictions of $R_S$ for the Solar System giant planets are listed in Table 1, together their actually observed dipole moments and present values of the solar wind properties.

As a giant planet ages, its internal heat source recedes and thus the planet's internal dynamo as well as its magnetic field weaken[21]. Yet, the stellar wind also weakens as the stellar activity decreases. The combination of both effects induces a net expansion of the magnetosphere over billions of years. Figure 10 shows the evolution of $R_S$ for a Jupiter-analog in the center of the Sun's HZ, following methods presented by Heller and Zuluaga (2013): $M_p = M_{Jup}$, the planetary core mass $M_{core} = 10$ $M_\oplus$, and the planet's rotation $P_{rot} = 10$ h. Magnetospheres of HZ giant planets could be very compressed during the first billion years, thereby leaving their moons unshielded. As an example, compare the blue spiral in Figure 10, which denotes $R_S$, with the orbits of the Galilean moons. While an Io-analog satellite would be coated by the planetary magnetic field as early as 500 Myr after formation, a moon in a Europa-wide orbit would be protected after about 2 Gyr, and a satellite in a Ganymede-wide orbit would only be shielded after 3.5 Gyr. This evolutionary effect of an exomoon's magnetic environment can thus impose considerable constraints on its habitability. Heller and Zuluaga (2013), while focussing on moons around giant planets in the HZ of K stars, concluded that exomoons beyond 20 planetary radii from their host will hardly ever be shielded by their planets' magnetospheres. Moons in orbits between 5 and 20 $R_p$ could be shielded after several hundred Myr or some Gyr, and they could also be habitable from an illumination and tidal point of view. Moons closer to their planets than 5 $R_p$ could be shielded during the most hazardous phase of the stellar wind, that is, during the first $\approx 100$ Myr, but they would likely not be habitable in the first place, due to enormous tidal heating and strong planetary illumination (Heller and Barnes, 2014).

---

[21] The evolution of the magnetic field strength could be much more complex if the interplay between the energy available energy for the dynamo and the Coriolis forces are taken into account (Zuluaga and Cuartas, 2012).

| | | | | | | | | Shielding | |
| --- | --- | --- | --- | --- | --- | --- | --- | --- | --- |
| **Planet** | $\mathcal{M}$ ($\mathcal{M}_\oplus$) | $R_S^{\text{obs}}$ ($R_p$) | $\bar{R}_S^{\text{th}}$ ($R_p$) | $\bar{R}_S^{\text{th,HZ}}$ ($R_p$) | **Moons** | $a_{ps}$ ($R_p$) | | **Actual** | **HZ** |
| Jupiter | 1800 | $45-80$ | $38-48$ | $22-24$ | Io | 5.9 | | CS | CS |
| | | | | | Europa | 9.4 | | CS | CS |
| | | | | | Ganymede | 15.0 | | CS | CS |
| | | | | | Callisto | 26.3 | | CS | **PS** |
| Saturn | 580 | $17-30$ | $17-23$ | $8.5-9.0$ | Enceladus | 4.0 | | CS | CS |
| | | | | | Titan | 20.3 | | **PS** | **US** |
| Uranus | 50 | 18 | $23-33$ | $8.9-9.3$ | Titania | 17 | | CS | **US** |
| | | | | | Oberon | 22.8 | | **PS** | **US** |
| Neptune | 24 | $23-26$ | $21-32$ | $7.2-7.6$ | Triton | 14 | | CS | **US** |

**Table 1**: Measured magnetic properties of giant planets in the Solar System. $\mathcal{M}$: Magnetic dipole moment (Bagenal, 1992; Guillot, 2005). $R_S^{\text{obs}}$: observed range of magnetospheric standoff radii (Arridge et al., 2006; Huddleston et al., 1998). Note that standoff distances vary with solar activity. $R_S^{\text{th}}$: range of predicted average standoff distances. Predicted standoff distances depend on magnetospheric compressibility (see text). $R_S^{\text{th,HZ}}$: extrapolation of $R_S^{\text{th}}$ assuming that the planet would be located at 1 AU from the Sun. Semi-major axes $a_{ps}$ of selected major moons are shown for comparison. The final two columns depict the shielding status (see Fig. 13) of the moons for their actual position in the Solar System and assuming a distance of 1 AU to the Sun.

### 5.3.1 Unshielded exomoons

The fraction of a moon's orbit spent inside the planet's magnetospheric cavity defines three different shielding conditions: (i.) unshielded (US), (ii.) partially shielded (PS), and (iii.) completely shielded (CS) (Fig. 9). Depending on its membership to any of these three classes, different phenomena will affect an exomoon, some of which will be conducive to habitability and others of which will threaten life.

Assuming that the dayside magnetosphere is approximately spherical and that the magnetotail is cylindrical, a moon orbiting its host planet at a distance of about 2 $R_S$ will spend more than 50% of its orbit outside the planetary magnetosphere. To develop habitable surface conditions, such a partially shielded exomoon would need to have an intrinsic magnetic field similar in strength to that required by a terrestrial planet at the same stellar distance (Zuluaga et al., 2013; Vidotto et al., 2013). For moons in the HZ of Sun-like stars, intrinsic magnetic dipole moments larger than about that of Earth would allow for habitable surface conditions. Moons near the HZs around less massive stars, however, which show enhanced magnetic activity, need intrinsic dipole moments that are several times larger to prevent an initial satellite atmosphere from being exposed to the strong stellar wind (Vidotto et al., 2013). As formation models predict that moons can hardly be as massive as Earth (see Section 3), the maximum magnetic field attainable by an hypothetical exomoon may be insufficient to prevent atmospheric erosion (Williams et al., 1997). Perhaps moons with alternative internal heat sources could drive a long-lived, sufficiently strong internal dynamo.

Ganymede may serve as an example to illustrate this peril. It is the only moon in the Solar System with a strong intrinsic magnetic field (Ness et al., 1979). With a dipole moment about $2 \times 10^{-3}$ times that of Earth, a Ganymede-like, unshielded moon (see "US" zone in Fig. 9) orbiting a giant planet in the HZ of a Sun-like star would only have a standoff radius of about 1.5 Ganymede radii at an age of 1 Gyr. This would expose a potentially thin and extended atmosphere of such a moon to the high-energy particles from the stellar wind, eventually stripping off the whole atmosphere.

### 5.3.2 Partially shielded exomoons

Inside the planetary magnetosphere, the main threat to a partially shielded exomoon is that of the so-called co-rotating magnetoplasma (Neubauer et al., 1984). Magnetospheric plasma, which is dragged by the planetary magnetic field to obtain the same rotational angular velocity as the planet, will produce a "magnetospheric wind". Around gas giants akin to those in the Solar System, the circumplanetary angular velocity of the co-rotating plasma is typically much higher than the Keplerian velocity of the moons. Hence, the magnetospheric wind will hit exomoon atmospheres and may erode them in orbits as wide as a significant fraction of the standoff distance. This effect could be as intense as effects of the direct stellar wind. For instance, it has been estimated that Titan, which is partially shielded by Saturn's magnetosphere, could have lost 10% of its present atmospheric content just because of charged particles from the planet's co-rotating plasma (Lammer and Bauer, 1993).

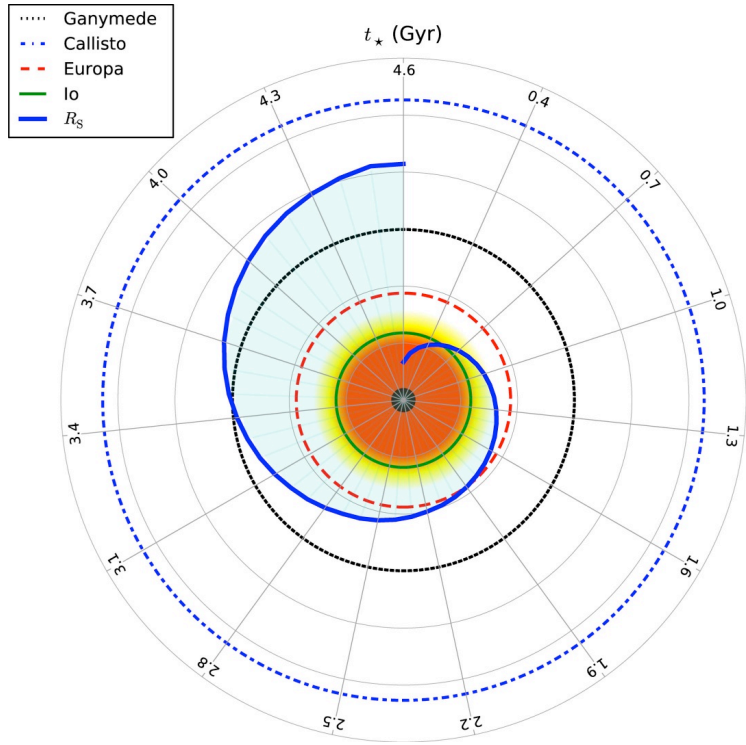

**Figure 10**: Evolution of the standoff radius $R_S$ (thick blue line) around a Jupiter-like planet in the center of the stellar HZ of a Sun-like star. For comparison, the orbits of the Galilean moons are shown (see legend). Angles with respect the vertical line encode time in Gyr. Time starts at the "12:00" position in 100 Myr and advances in steps of 0.3 Gyr up to the present age of the Solar System. Thin gray circles denote distances in intervals of 5 planetary radii. The filled circle in the center denotes the planetary radius.

Towards the inner edge of the HZ around low-mass stars, the stellar wind flux is extremely strong. What is more, the planetary magnetosphere will be small, thereby increasing the density of the magnetospheric wind. Other sources of plasma such as stellar wind particles, planetary ionosphere gasses, and ions stripped off from other moons could also be greatly enhanced.

### 5.3.3 Completely shielded exomoons

Moons that are completely shielded by the planetary magnetic shield are neither exposed to stellar wind nor to significant amounts of cosmic high-energy particles. What is more, being well inside the magnetosphere, the co-rotating plasma will flow at velocities comparable to the orbital velocity of the moon. Hence, no strong interactions with the satellite atmosphere are expected.

On the downside, planetary magnetic fields can trap stellar energetic particles as well as cosmic rays with extremely high energies. Around Saturn and Jupiter, electrons, protons, and heavy ions with energies up to hundreds of GeV populate the inner planetary magnetospheres within about ten planetary radii, thereby creating what is known as radiation belts. Around Jupiter, as an example, the flux of multi-MeV electrons and protons at a distance of about 15 $R_{Jup}$ to the planet is between $10^6$ and $10^8$ cm$^{-2}$ s$^{-1}$ (Divine and Garrett, 1983), that is four to six orders of magnitude larger than that of solar high-energy particles received by Earth. Exposed to these levels of ionizing radiation, moon surfaces could absorb about 1 to 10 J kg$^{-1}$ min[-1] (National Research Council, 2000)[22], which is many orders of magnitude above the maximum tolerable radiation levels for most microorganisms found on Earth (Baumstark-Khan and Facius, 2002).

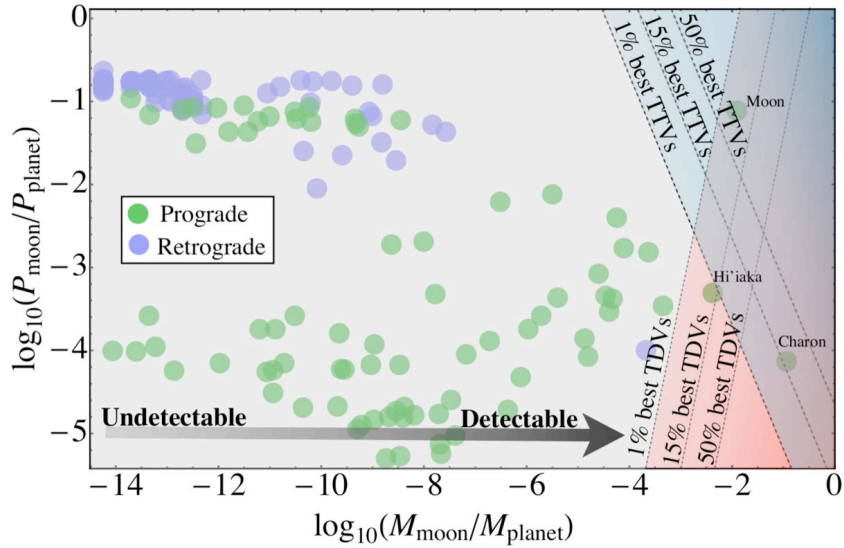

**Figure 11**: Period-ratio versus mass-ratio scatter plot of the Solar System moons. Transit timing and duration variations (TTV and TDV) exhibit complementary sensitivities with the period-ratio. Using the *Kepler* timing measurements from Ford et al. (2012), one can see that the tip of the observed distribution is detectable. The above assumes a planetary period of 100 days and a baseline of 4.35 years of *Kepler* data.

Magnetic field intensities inside compressed magnetospheres will be larger for a given dipole moment of a Solar System giant planet. Moreover, stellar high-energy particle fluxes, feeding radiation belts, will be also larger. As a result, the flux of high energy particles accelerated inside the magnetosphere of giant planets in the HZ could be strongly enhanced with respect to the already deadly levels expected on Jupiter's moon Europa.

## 6. Detection Methods for Extrasolar Moons

Within the arena of transiting planets, there are two basic effects which may betray the presence of an exomoon: (i.) dynamical effects which reveal the mass ratio between the satellite and the planet ($M_s/M_p$), and (ii.) eclipse effects which reveal the radius ratio between the satellite and the star ($R_s/R_\star$). Detecting both effects allows for a measurement of the bulk density and thus allows one to distinguish between, say, icy moons versus rocky moons.

Beyond the possibility to detect extrasolar satellites when using the stellar transits of a planet-moon system, several other methods have been proposed over the past decade**.** Cabrera and Schneider (2007) suggested that an exomoon might induce a wobble of a directly imaged planet's photocenter and that planet-moon eclipses might be detectable for directly imaged planets (see also Sato and Asada 2010; Pál 2012). Excess emission of transiting giant exoplanets in the spectrum between 1 and 4 μm (Williams and Knacke 2004) or enhanced infrared of terrestrial planets (Moskovitz et al. 2009; Robinson 2011) might also indicate the presence of a satellite. Further approaches consider the Rossiter-McLaughlin effect (Simon et al. 2010; Zhuang et al. 2012), pulsar timing variations (Lewis et al. 2008), microlensing (Han and Han 2002), modulations of a giant planet's radio emission (Noyola et al. 2014), and the the generation of plasma tori around giant planets caused by the tidal activity of a moon (Ben-Jaffel and Ballester 2014). In particular, the upcoming launch of the James Webb Space Telescope (*JWST*) inspired Peters and Turner (2013) to propose the possibility of detecting an exomoon's thermal emission. In the following, we discuss the dynamical effects that may reveal an exomoon as well as the prospects of direct photometric transit observations and constraints imposed by white and red noise. Possibilities of detecting extremely tidally heated exomoons via their thermal emission will also be illustrated.

*6.1 Dynamical effects on transiting exoplanets*

---

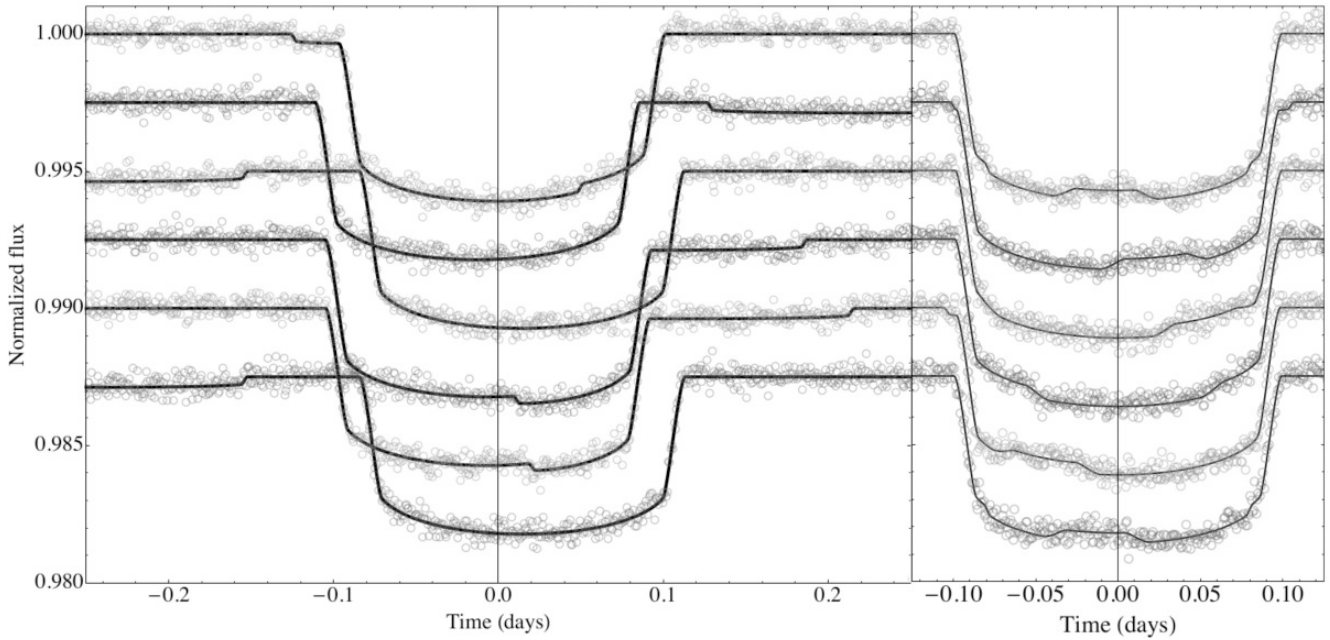

**Figure 12**: (*Left*) Six simulated transits using LUNA (Kipping, [2011b](#)) of a HZ Neptune around an M2 star with an Earth-like moon on a wide orbit (90% of the Hill radius). The moon can be seen to exhibit "auxiliary transits" and induce TTVs. (*Right*) Same as left, except the moon is now on a close-in orbit (5% of the Hill radius), causing "mutual events". Both plots show typical *Kepler* noise properties for a 12th-magnitude star observed in short-cadence.

The first technique ever proposed to detect an exomoon comes from Sartoretti and Schneider ([1999](#)), which falls into the dynamical category and is now often referred to as transit timing variations, or simply TTV, although it was not referred to as this in the original paper. For a single moon system, the host planet and the companion orbit a common barycenter, which itself orbits the host star on a Keplerian orbit. It therefore follows that the planet does not orbit the star on a Keplerian orbit and will sometimes transit slightly earlier or later depending upon the phase of the moon. These deviations scale proportional to $P_P \times (M_s/M_P) \times (a_s/a_P)$ and can range from a few seconds to exhibit up to a few hours for terrestrial moons. The original model of Sartoretti and Schneider ([1999](#)) deals with circular, coplanar moons but extensions to non-coplanar and eccentric satellites have been proposed since (Kipping, [2009a](#), [2011a](#)). Three difficulties with employing TTV in isolation are that, firstly, other effects can be responsible (notably perturbing planets); secondly, the TTV waveform is guaranteed to be below the Nyquist frequency and thus undersampled, making a unique determination of $M_s$ untenable; and, thirdly, moons in close orbits blur their planet's TTV signature as the planet receives a substantial tangential acceleration during each transit (Kipping [2011a](#)).

TTV can be thought of as being conceptually analogous to the astrometric method of finding planets, since it concerns changes in a primary's position due to the gravitational interaction of a secondary. Astrometry, however, is not the only dynamical method of detecting planets. Notably, Doppler spectroscopy of the host star to measure radial velocities has emerged as one of the work horses of exoplanet detection in the past two decades. Just as with astrometry, an exomoon analogy can be devised by measuring changes in a planet's transit duration, as a proxy for its velocity (Kipping, [2009a](#)). Technically though, one is observing tangential velocity variations rather than those in the radial direction. Transit duration variations due to velocity variations, dubbed TDV-V, scale as $(P_P/P_s) \times (M_s/M_P) \times (a_s/a_P)$ and vary from seconds to tens of minutes in amplitude for terrestrial satellites.

Just as radial velocity is complementary to astrometry, TDV-V and TTV are complementary since TTV is more sensitive to wide-orbit moons (sensitivity scales as $\propto a_s$) and TDV is more sensitive to close-orbit moons (sensitivity scales as $\propto a_s^{-1/2}$). Furthermore, should one detect both signals, the ratio of their root-mean-square amplitudes (that is, their statistical scatter) yields a direct measurement of $P_s$, which can be seen via inspection of the aforementioned scalings. This provides a powerful way of measuring $P_s$ despite the fact that the signals are undersampled. Finally, TDV-V leads TTV by a $\pi/2$ phase shift in amplitude (Kipping [2009a](#)) providing a unique exomoon signature, which even for undersampled data can be detected with cross-correlation techniques (Awiphan and Kerins, [2013](#)). Just as with TTV though, a TDV-V signal in isolation suffers from both model and parameter degeneracies.

An additional source of transit duration variations comes from non-coplanar satellite systems, where the planet's reflex motion causes position changes not only parallel to the transit chord (giving TTV) but also perpendicular to it. These variations cause the apparent impact parameter of the transit to vary leading to transit impact parameter induced transit duration variations, known as TDV-TIP (Kipping, 2009b). This is generally a small effect of order of seconds but can become much larger for grazing or near-grazing geometries. Should the effect be detected, it can be shown to induce a symmetry breaking that reveals whether a moon is prograde or retrograde in its orbit, thereby providing insights into the history and formation of the satellite.

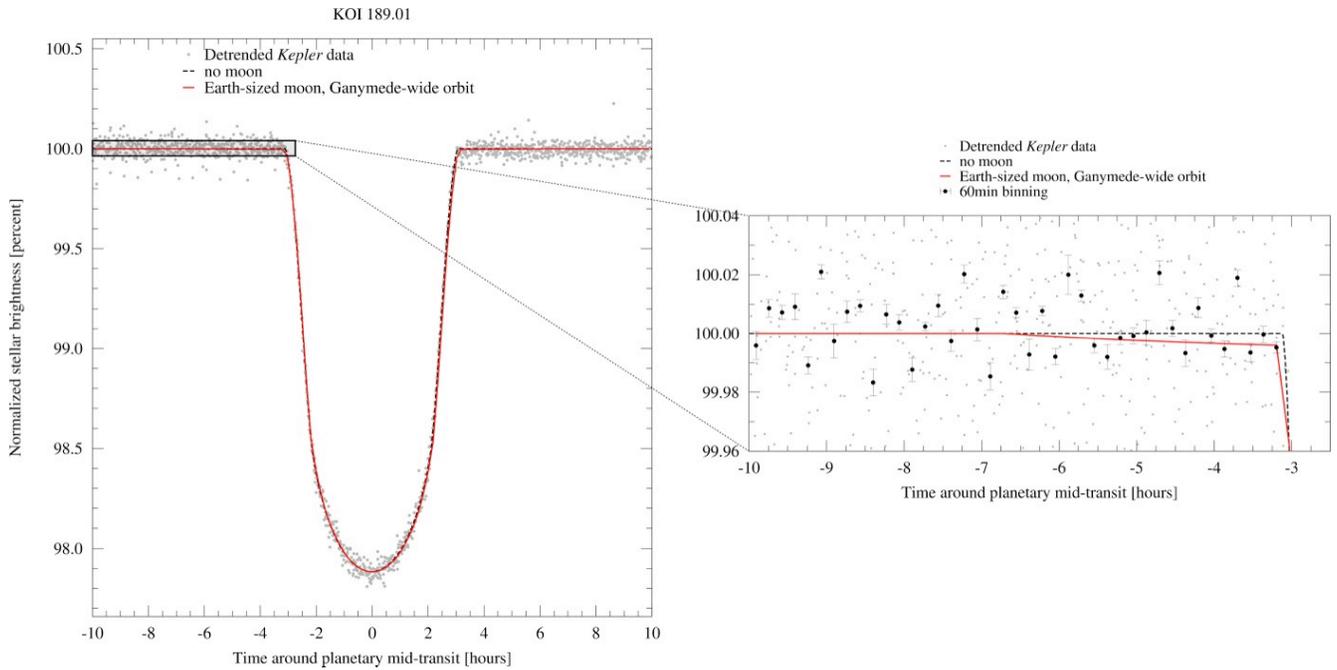

**Figure 13**: Transit of the almost Jupiter-sized planet candidate KOI189.01 around a star of about 0.7 solar radii. Gray dots indicate the original phase-folded *Kepler* light curve, the black dashed line indicates a model for the transit assuming a planet only, and the red line assumes an Earth-sized moon in an orbit that is 15 planetary radii wide. Black dots in the right panel indicate data that is binned to 60 minutes. The photometric orbital sampling effect appears in the right panel as a deviation between the red solid and the black dashed line about 6.5 hours before the planetary mid-transit.

For all of the aforementioned techniques, multiple moons act to cancel out the overall timing deviations, except for resonant cases. In any case, the limited number of observables (two or three) make the detection of multiple moons untenable with timing data alone. Despite this, there are reasons to be optimistic for the detection of a large quasi-binary type exomoon. Kipping et al. (2009) and Awiphan and Kerins (2013) estimated that the *Kepler* space telescope is sensitive to telluric moons. Recent statistics on the timing sensitivity of *Kepler* by Ford et al. (2012) reveal that the tip of the known population of moons in our own Solar System would be detectable with *Kepler* too, as shown in Fig. 11.

In terms of detectable habitable exomoons, Kipping et al. (2009) took into account orbital stability criteria from Barnes and O'Brien (2002) and found that moons more massive than about 0.2 $M_\oplus$ and orbiting a Saturn-mass planet in the center of the stellar HZ can be detectable with *Kepler*-class photometry if the star is more massive than about 0.3 $M_\odot$ and has a *Kepler* magnitude ≤ 12 (Fig. 6 of Kipping et al, 2009). More massive stars would need to be brighter because the planetary transit and its TTV and TDV signals become relatively dimmer. Simulations by Awiphan and Kerins (2013) suggest that moons in the habitable zone of a 12.5 *Kepler* magnitude M dwarf with 0.5 times the mass of the Sun would need to be as massive as 10 $M_\oplus$ and to orbit a planet less massive than about 1/4 the mass of Saturn to be detectable via their planet's TTV and TDV. Such a system would commonly be referred to as a planetary binary system rather than a planet-moon binary because the common center of gravity would be outside the primary's radius. From this result and similar results (Lewis, 2011b), the detection of a HZ moon in the ≈ 1400 d of *Kepler* data is highly challenging.

### 6.2 Direct eclipse effects

While dynamical effects of the planet can yield the system's orbital configuration, a transit of the moon itself not only provides a measurement of its radius but also has the potential to distort the transit profile shape leading to the erroneous derivation of TTVs and TDVs. Modeling the moon's own transit signal is therefore both a critical and inextricable component of hunting for exomoons.

Exomoons imprint their signal on the stellar light curve via two possible effects. The first is that the moon, with a wide sky-projected separation from the planet, transits the star and causes a familiar transit shape. These "auxiliary transits" can occur anywhere from approximately 93% of the Hill radius away from the planet (Domingos et al., 2006) to being ostensibly on-top of the planet's own transit (see Fig. 12 for examples). Since the phase of the moon will be unique for each transit epoch, the position and duration of these auxiliary transits will vary, which can be thought of as TTV and TDV of the moon itself, magnified by a factor of $M_p/M_s$. The position and duration variations of the moon will also be in perfect anti-phase with the TTVs and TDVs of the planet, providing a powerful confirmation tool.

The second eclipse effect caused by exomoons are so-called "mutual events". This is where the moon and planet appear separated at the start of a transit but then the moon eclipses (either in front or behind) the planet at some point during the full transit duration. This triple-overlap means that the amount of light being blocked from the star actually decreases during the mutual event, leading to what appears to be an inverted-transit signal (Fig. 12). Mutual events can occur between two planets too (Ragozzine and Holman, 2010), but the probability is considerably higher for a moon. The probability of a mutual event scales as $\approx R_p/a_s$ and thus is highest for close-in moons. One major source of false-positives here are starspots crossings (Rabus et al., 2009), which appear almost identical, but of course do not follow Keplerian motion.

Accounting for all of these eclipse effects plus all of the timing effects is arguably required to mount a cogent and expansive search for exomoons. Self-consistent modeling of these phenomena may be achieved with full "photodynamic" modeling, where disc-integrated fluxes are computed at each time stamp for positions computed from a three-body (or more) dynamical model. Varying approaches to moon-centric photodynamical modeling exist in the literature including a re-defined TTV for which the photocenter (Simon et al., 2007), circular/coplanar photodynamical modeling (Sato and Asada, 2009; Tusnski and Valio, 2011), and full three-dimensional photodynamic modeling (Kipping, 2011b) have been used.

Another way of detecting an exomoon's direct transit signature is by folding all available transits of a given system with the circumstellar orbital period into a single, phase-folded light curve. If more than a few dozen transits are available, then the satellites will show their individual transit imprints (Heller 2014). Each moon causes an additional transit dip before and after the planetary transit and thereby allows measurements of its radius and planetary distance. This so-called orbital sampling effect loses the temporal information content exploited in rigorous photodynamical modeling, meaning that the satellite's period cannot be determined directly. And without the satellite period, one cannot compute the planetary density via the trick described by Kipping (2010). Despite this, the computational efficiency of the OSE method makes it attractive as a quick method for identifying exomoon candidate systems, perhaps in the readily available *Kepler* data or the upcoming *Plato* space mission. But for such a detection, dozens of transits in front of a photometrically quiet M or K dwarf are required (Heller 2014). What is more, to ultimately confirm the presence of an exomoon, additional evidence would be required to exclude photometric variations due other phenomena, both astrophysical (e.g., stellar activity, rings) and instrumental (e.g., red noise). A confirmation could be achieved with photodynamical modeling as applied by the HEK team. As an example for the photometric OSE, Figure 13 presents the phase-folded Kepler light curve (gray dots) of the almost Jupiter-sized planetary candidate KOI189.01, transiting a $\approx 0.7\ R_\odot$ K star. Note how the "no moon" model (black dashed line) and the model for an Earth-sized moon (red solid line) diverge in the right panel! The latter is based on an improved model of Heller (2014), now including stellar limb darkening, amongst others, and is supposed to serve as a qualitative illustration but not as a statistical fit to the Kepler data for the purpose of this paper.

### 6.3 Photometric noise

One major factor constraining the detection of moons of transiting planets is the type of photometric noise contaminating the light curve.[23] A first detailed analysis of noise effects on the detection of exomoons with *Kepler* was performed by Kipping et al. (2009), showing that photon noise and instrument noise strongly increase for stellar apparent magnitudes $\gtrsim 13$. Combing shot noise, *Kepler*'s instrumental noise, and stellar variability with arguments from orbital stability, they obtained a lower limit of about 0.2 $M_\oplus$ for the detection of moons orbiting Saturn-like planets in the stellar HZ of bright M stars (with *Kepler* magnitudes < 11) by using TTV and TDV. This mass detection limit increases for Neptune- and Jupiter-like planets because of these planets' higher densities (see Fig. 3 in Kipping et al., 2009). What is more, absolute mass detection limits are hard to generalize because the TTV-TDV combined method constrains the planet-to-satellite mass ratio $M_s/M_p$. In their targeted search for exomoons, Kipping et al. (2013a) achieved accuracies as good as $M_s/M_p \approx 4\%$.

Lewis (2011a) investigated the effects of filtering on Sun-like stellar noise for a variant of TTV, photometric transit timing, and found that realistic photometric noise suppresses the detection of moons on wide circumplanetary orbits. In another work, Lewis (2013) confirmed that these noise sources also hamper the detection of moons around planets that follow distant circumstellar orbits. As a result, exomoons hidden in the noise of the *Kepler* data will need to have radii $\gtrsim 0.75\ R_\oplus$ to be detectable by direct eclipse effects. This value is comparable to the results of Simon et al. (2012), who found that moons of close-in planets, that is, with circumstellar orbital periods ≲ 10 days, could be detected by using their scatter peak method on *Kepler* short cadence data if the moons have radii ≳ 0.7 $R_\oplus$.

---

[23] In the case of moon detection through perturbation of the Rossiter-McLaughlin effect (Simon et al., 2010; Zhuang et al., 2012) one needs to consider types of noise that contaminate radial velocity measurements.

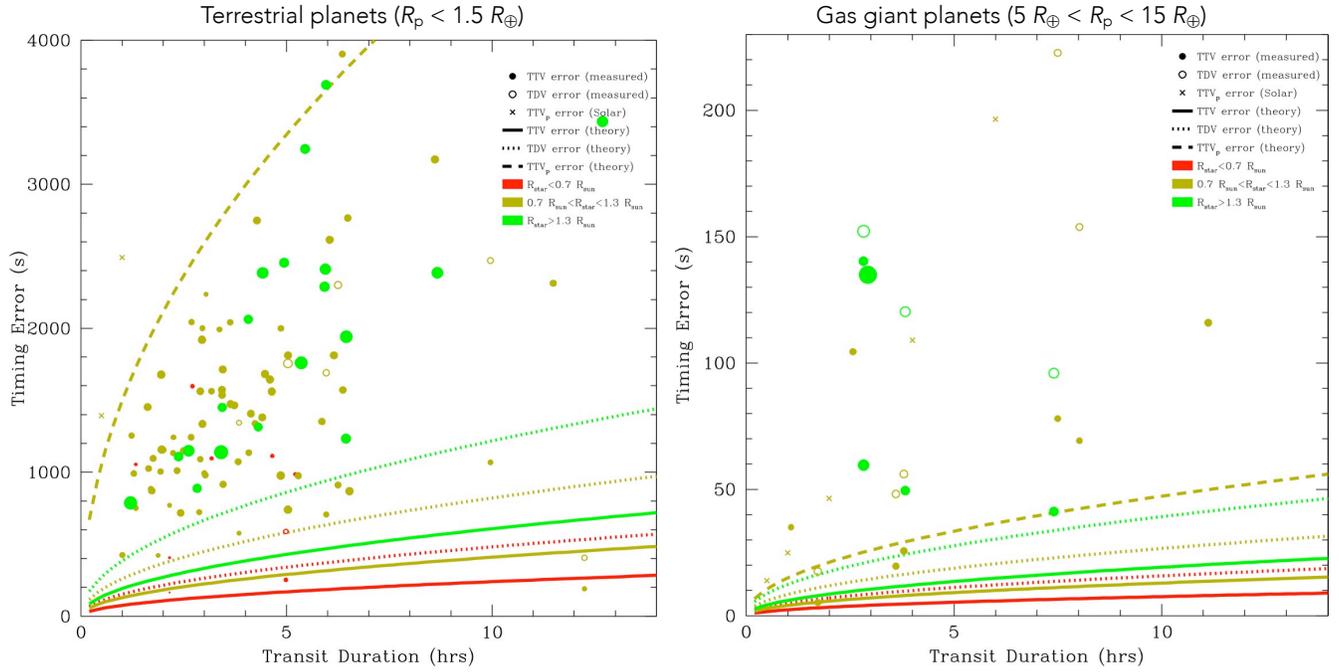

**Figure 14**: Calculated timing errors of the transit mid-time (filled circles) and duration (open circles) for *Kepler* planet candidates around stars with magnitudes < 13.5 (Mazeh et al., 2013). Planetary radii are estimated by assuming an Earth-like composition (left panel) or a mostly gaseous composition (right panel). Point size is proportional to the stellar radius, with color indicating if the radius is smaller than 0.7 solar radii ($R_\odot$) (red), between 0.7 and 1.3 $R_\odot$ (gold), or larger than 1.3 $R_\odot$ (green). Predictions of the errors in transit mid-time (solid line) and duration (dotted line) following Carter et al. (2008) assume a central transit and white noise for the case of a 1 $R_\oplus$ or 10 $R_\oplus$ radius planet orbiting a 0.7 $R_\odot$ (red), $R_\odot$ (gold), or 1.3 $R_\odot$ (green) star. Photometric transit timing errors, dominated by white noise (dashed line) and realistic solar-type noise (crosses) assume a 12th magnitude G dwarf and relative photometric precision of $2 \times 10^{-5}$ in a 6.5 hour exposure (Lewis, 2013).

Uncorrelated noise that follows a Gaussian probability distribution

$$f(x) = \frac{1}{\sigma\sqrt{2\pi}} e^{\frac{(x-\mu)^2}{2\sigma^2}},$$

(6)

where $\sigma$ is the standard deviation, $x$ is the perturbation to the light curve due to noise, and $\mu$ is the mean of the noise (usually zero), is called "white noise", and it makes moon detection relatively straightforward. At a given time, $f(t)$ is assumed to be uncorrelated with the past. As a result, white noise contains equal power at all spectral frequencies. White noise or noise that is nearly white is produced by many instrumental or physical effects, such as shot noise.

Processes in stars, in Earth's atmosphere, and in telescopes can lead to long-term correlated trends in photometry that show an overabundance of long-wavelength components. In astrophysics, these effects are collectively referred to as red noise (Carter and Winn, 2009). Stars pulsate, convect, show spots and rotate, each of which leaves a distinctive red noise signature on the light curve (Aigrain et al., 2004). Solar oscillations have a typical amplitude of a few parts per million and period on the order of five minutes, while for other stars the specific details depend on the stellar size and structure (Christensen-Dalsgaard, 2004). In Sun-like stars, stellar convection leads to granulation, which produces red noise with characteristic timescales between minutes and hours. Finally, starspots can cause red noise in transit light curves, first through spot crossing events and second through spot evolution or rotational modulation. If a planet passes in front of a starspot group during transit, a temporary short-term increase in brightness results. This effect was predicted (Silva, 2003) and has finally been observed in many transiting systems (Pont et al., 2007; Sanchis-Ojeda and Winn, 2011; Kundurthy et al., 2011). Evolution and rotational modulation of starspots can cause long-term photometric variation over tens of days. This variation has been observed and modeled for a range of stars including CoRoT-2a (Lanza et al., 2009a) and CoRoT-4a (Lanza et al.,

2009b).

Ground-based observations can be substantially contaminated with red noise induced by Earth's atmosphere (Pont et al., 2006), for example, by the steadily changing value of the air mass over the course of a night. The operation of the telescope, as well as the telescope surroundings, can also introduce red noise, as has been noted by the ground-based WASP (Street et al., 2004) and HATNet surveys (Bakos et al., 2002, 2009). For the space-based *Kepler* and *CoRoT* missions, a number of different processes can lead to trends and jumps in the data, including effects from spacecraft motion, motion of the image on the CCD, and cosmic rays (Auvergne et al., 2009; *Kepler Characteristics Handbook*[24]). When a spacecraft rotates to reposition solar panels or to downlink data, or when it passes into Earth's shadow[25], the varying solar irradiation on various parts of the spacecraft's surface can cause transient photometric perturbations. In addition, the target's image can move on the CCD, and, finally, damage due to cosmic rays can lead to degradation of the CCDs. Cosmic rays can be isolated as well as certain specific events such as the passage of the south atlantic anomaly in the case of *CoRoT*, or the three large coronal mass ejections in *Kepler's* quarter 12. While processing pipelines work to reduce effects from cosmic rays (*Kepler Data Processing Handbook*[26]), they are never completely removed in all cases.

### *6.3.1 Effects of red noise on searches for exomoons*

While uncertainties in the transit mid-time and transit duration are dominated by errors in the ingress and egress parts of the light curve, relatively short lengths of time, red noise acts over longer periods. Hence, these techniques should be quite robust to red noise. However, transit mid-time (filled circles in Fig. 14) and duration (open circles) errors (Mazeh et al., 2013) of *Kepler* Objects of Interest are up to many factors above the values predicted by assuming white noise only. While inclination, stellar luminosity, stellar radius, and planetary radius can explain some of this discrepancy, some of it is undoubtedly due to red noise.

Carter and Winn (2009) investigated the effect of red noise on both mid-time and duration errors with a power spectrum following a power law, showing that parameters can be effectively recovered through use of a wavelet transform. Alternatively, Kipping et al. (2009) and later Awiphan and Kerins (2013) modeled red noise as white noise plus a set of longer period sinusoids. While Kipping et al. (2009) ended up neglecting the difference between white and red noise, their simulations show that the distribution of transit mid-times became slightly non-gaussian, in particular, the tails of the distribution became fatter (Fig. 2 of Kipping et al., 2009). Awiphan and Kerins (2013) confirmed this effect and concluded that this variety of red noise decreases moon detectability. Other recent studies suggest that the presence of starspots, especially near the stellar limb, can alter transit durations and timings (Silva-Válio, 2010; Barros et al., 2013), reducing sensitivity for spotted stars. Mazeh et al. (2013) confirmed this trend in transit timing errors for numerous Kepler Objects of Interest. In particular, they found that TTVs correlate with stellar rotation in active stars, indicating again that noise components due to starspots could alter timing results.

In addition to the standard transit timing technique, photometric transit timing has been proposed (Szabó et al., 2006), a technique that measures timing perturbations with respect to the average transit time weighted by the dip depth. Using solar photometric noise from the Solar and Heliosphere Observatory as a proxy, Lewis (2013) found that realistic red stellar photometric noise dramatically degraded moon detectability for this technique (see crosses in Fig. 14), compared to white noise for the case where the transit duration was longer than three hours. What is more, this degradation was not completely reversed by filtering, independent of the filtering method. This result is unfortunate, as the photometric transit timing statistic is proportional to the moon radius squared (Simon et al., 2007) as opposed to the moon's mass (Sartoretti and Schneider, 1999), but unsurprising as photometric transit timing uses data from both within and outside the planetary transit, compared to transit timing variation, which only uses shorter sections of data.

While Tusnski and Valio (2011) assumed white photometric noise, other projects have started to consider the effect of red noise in their analyses. The "Hunt for Exomoons with Kepler" (HEK) (Kipping et al., 2012) addresses the effect of red noise in a number of ways. First, they reject candidates with high levels of correlated noise (Kipping et al., 2013a). Second, a high-pass cosine filter is used to remove long-term trends from the data. The possibility of modeling the effects of starspots has also been suggested (Kipping, 2012). To help distinguish between star spot crossings and mutual events, rules have been proposed, for example, it is required that a mutual event have a flat top (Kipping et al., 2012). Also, to allow for secure detection, it is required that a given moon is detected by several detection methods. In an attempt to quantify the true error on detected moon properties, fake moon transits were injected into real data for the case of Kepler-22 b (Kipping et al., 2013b), indicating that an Earth-like moon, if present, would have been detected.

---

[24] http://archive.stsci.edu/kepler/manuals/Data_Characteristics.pdf

[25] *Kepler* orbits the Sun in an Earth-trailing orbit, where it avoids transits in the shadow of Earth. At an altitude of 827 km, *CoRoT* is in a polar orbit around the Earth, and occasionally passes the Earth's shadow.

[26] http://archive.stsci.edu/kepler/manuals/KSCI-19081-001_Data_Processing_Handbook.pdf

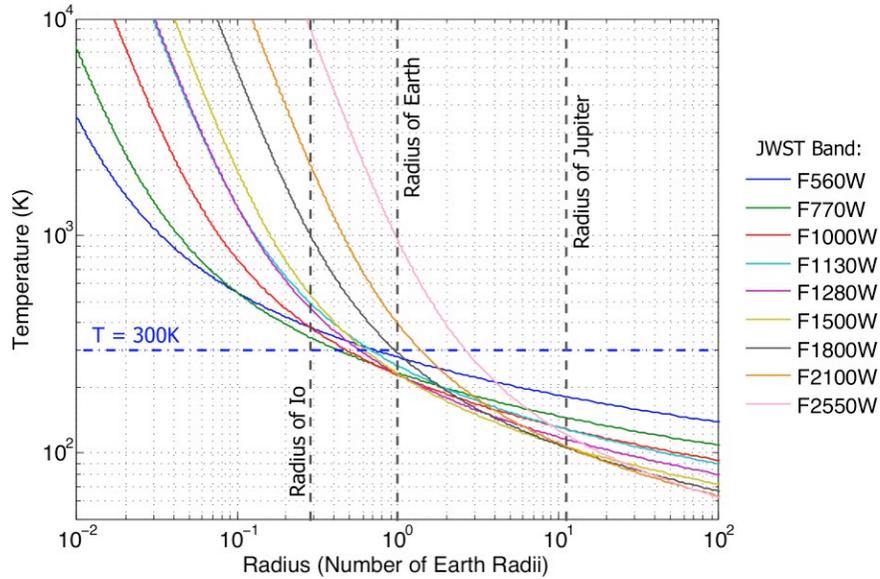

**Figure 15**: 5σ detection limits of JWST's MIRI for tidally heated exomoons with radii along the abscissa and effective temperatures along the ordinate (following Turner and Peters, 2013). 10,000s of integration are assumed for a star 3 pc from the Sun. MIRI's nine imaging bands are indicated with different line colors, their names encoding the wavelength in units of microns times 100. Dashed vertical black lines denote the radii of Io, Earth, and Jupiter, which is roughly equal to that of a typical brown dwarf.

### *6.4 Direct imaging of extrasolar moons*

#### *6.4.1 Principles of direct imaging*

Direct imaging of exoplanets, especially those in the stellar HZ, is extremely difficult because of the very small star-planet angular separation and the high contrast ratio between the star and planet. In fact, all exoplanets that have been directly imaged are well separated from their host star, and are young systems that are still hot from formation, rather than being heated by stellar irradiation, with effective temperatures around 1000 K. Examples of directly imaged exoplanets include the HR8799 planets, β Pic b, LkCa15b, κ And b, and GJ 504b (Marois et al., 2008; Lagrange et al., 2008; Kraus and Ireland, 2011; Carson et al., 2013; Kuzuhara et al., 2013). Intuitively, one would expect exomoons to be even more difficult to be imaged directly. For exomoons similar to those found in the Solar System, this is likely the case. However, satellites that are heated by sources other than stellar irradiation, such as tidal heating, can behave completely differently.

There has been considerable discussion in the literature of the existence of tidally heated exomoons (THEMs) that are extrasolar analogs to Solar System objects such as Io, Europa, and Enceladus (Peale et al., 1979; Yoder and Peale, 1981; Ross and Schubert, 1987; Ross and Schubert, 1989; Nimmo et al., 2007; Heller and Armstrong 2014). Yet, the possibility of imaging the thermal emission from unresolved THEMs was proposed only recently by Peters and Turner (2013)[27]. From an observational point of view, direct imaging of exomoons has several advantages over exoplanet direct imaging. THEMs may remain hot and luminous for timescales of order the stellar main sequence lifetime and thus could be visible around both young and old stars. Additionally, THEMs can be quite hot even if they receive negligible stellar irradiation, and therefore they may be luminous even at large separations from the system primary. This will reduce or even eliminate the inner-working angle requirement associated with exoplanet high contrast imaging. Furthermore, tidal heating depends so strongly on orbital and physical parameters of the THEM that plausible systems with properties not very different from those occurring in the Solar System will result in terrestrial planet sized objects with effective temperatures up to 1000 K or even higher in extreme but physically permissible cases. The total luminosity of a THEM due to tidal heating is given by

$$L_{\text{tid}} = \left( \frac{6272\pi^7 G^5}{9747} \right)^{1/2} \left( \frac{R_s^7 \rho_s^{9/2}}{\mu_s Q_s} \right) \left( \frac{e_{ps}^2}{\beta_{ps}^{15/2}} \right) , \tag{7}$$

where $\mu_s$ is the moon's elastic rigidity, $Q_s$ is the moon's tidal dissipation function (or quality factor), $e_{ps}$ is the eccentricity of the planet-satellite orbit, $\beta_{ps}$ is the satellite's orbital semi-major axis in units of Roche radii, $\rho_s$ is the satellite density, and $R_s$ is the satellite radius (Scharf, 2006; Peters and Turner, 2013). Equation (7) is based on the tidal heating equations originally derived by Reynolds et al. (1987) and Segatz et al. (1988) and assumes zero obliquity. The terms that depend on the exomoon's physical properties and those that describe its orbit are grouped separately. Although $\beta_{ps}$ is grouped with the orbital terms, in addition to its linear dependence on the moon's semi-major axis $a_{ps}$, it also scales with the planetary mass and the satellite density as $(M_p/\rho_s)^{1/3}$ for a fixed $a_{ps}$.

---

[27] For exoplanets, direct imaging refers to the imaging of a planet that is spatially resolved from its host star. In the model proposed by Peters and Turner (2013), the tidally heated moon is not spatially resolved from its host planet, but it rather adds to the thermal flux detected at the circumstellar orbital position of the planet. Despite this unresolved imaging, we here refer to this exomoon detection method as direct imaging.

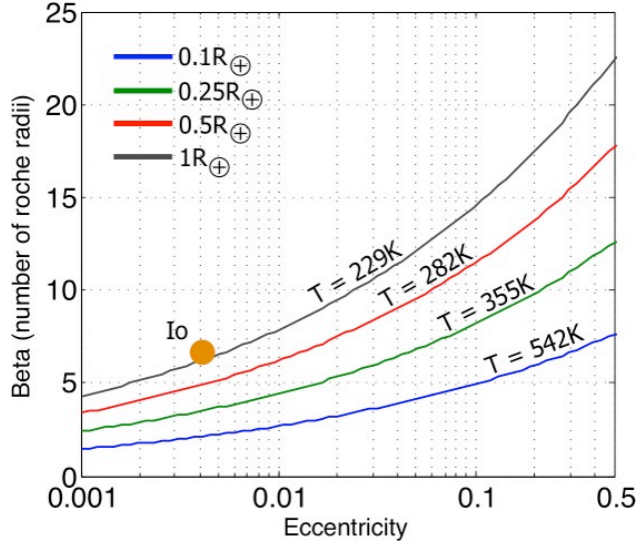

**Figure 16**: Minimum detectable eccentricity (abscissa) and semi-major axis in units of Roche radii ($\beta_{ps}$, ordinate) for tidally-heated exomoons between 0.1 and 1 $R_\oplus$ in size (Turner and Peters, 2013). The assumed bodily characteristics are described around Eqs. (5) - (6). The orange dot indicates Io's eccentricity and $\beta_{ps}$, but note that its size is 0.066 $R_\oplus$. Satellites below their respective curve will be detectable by MIRI as far as 3 pc from Earth, provided they are sufficiently separated from their star.

Following Peters and Turner (2013), the scaling relation for Eq. (7) relative to the luminosity of Earth ($L_\oplus = 1.75 \times 10^{24}$ ergs/s) is

$$L_s \approx L_\oplus \left[ \left( \frac{R_s}{R_\oplus} \right)^7 \left( \frac{\rho_s}{\rho_\oplus} \right)^{\frac{9}{2}} \left( \frac{36}{Q_s} \cdot \frac{10^{11} \frac{\text{dynes}}{\text{cm}^2}}{\mu_s} \right) \right] \times \left[ \left( \frac{e_{ps}}{0.0028} \right)^2 \left( \frac{\beta_{ps}}{8} \right)^{-\frac{15}{2}} \right] . \tag{8}$$

Note that Eq. (8) adopts $Q_s = 36$, $\mu_s = 10^{11}$ dynes/cm$^2$ as for Io (Peale et al., 1979; Segatz et al., 1988), and Earth's radius and density as reference values. The reference values of $\beta_{ps}$ and $e_{ps}$ were then chosen to give $L_\oplus$. If we assume that THEMs are blackbodies, we can use the exomoon's luminosity to approximate its effective temperature. The blackbody assumption is a reasonable approximation, but in general we would expect THEMs to emit excess light at bluer wavelengths due to hotspots on the surface. Additionally, exomoons with atmospheres are likely to have absorption lines in their spectrum. Assuming a blackbody, we can define a satellite's effective temperature ($T_s$) from the luminosity via the Stefan-Boltzmann law as a scaling relation relative to a 279 K exomoon, this temperature corresponding to the equilibrium temperature of Earth:

$$T_s \approx 279K \left[ \left( \frac{R_s}{R_\oplus} \right)^{\frac{5}{4}} \left( \frac{\rho_s}{\rho_\oplus} \right)^{\frac{9}{8}} \left( \frac{36}{Q_s} \cdot \frac{10^{11} \frac{\text{dynes}}{\text{cm}^2}}{\mu_s} \right)^{1/4} \right] \times \left[ \left( \frac{e_{ps}}{0.0028} \right)^{\frac{1}{2}} \left( \frac{\beta_{ps}}{8} \right)^{-\frac{15}{8}} \right] \tag{9}$$

(Peters and Turner, 2013). As an example, Eq. (9) yields $\approx$ 60 K for Io, corresponding to the effective temperature of Io if it were not irradiated by the Sun. Note that Eq. (9) adopts the same reference values as Eq. (8) and that it assumes only tidal heating with no additional energy sources, such as stellar irradiation or interior radiogenic heat.

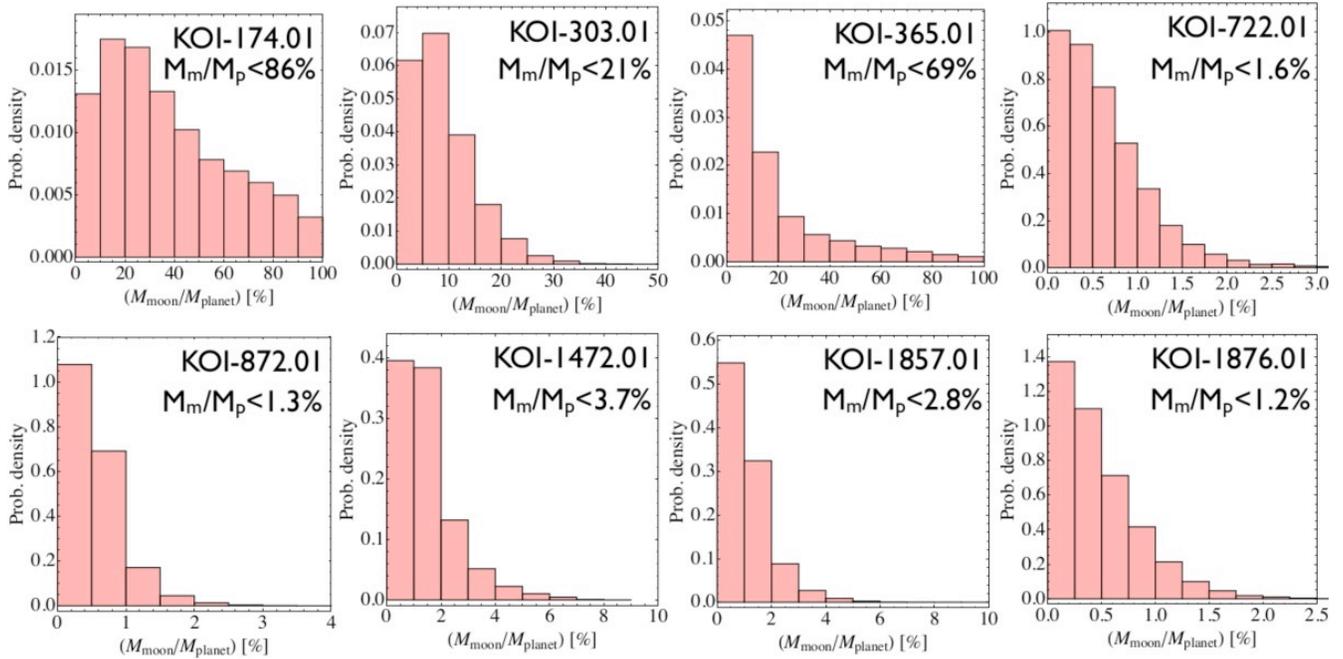

**Figure 17**: Moon-to-planet mass ratio constraints derived so far by the HEK project (Nesvorný et al., 2012; Kipping et al., 2013a). Kepler-22b has also been studied and yields $M_s < 0.5\ M_\oplus$ but is not shown here as the paper is still under review at the time of writing.

An exomoon's brightness is expected to be variable for multiple reasons, including eclipses of the exomoon behind its host planet (Heller, 2012), phase curve variations due to temperature variation across an object's surface (Heller and Barnes, 2013), and volcanism.

### 6.4.2 Direct imaging with future facilities

If THEMs exist and are common, are they detectable with current and future instrumentation? As shown by Peters and Turner (2013), Spitzer's IRAC could detect an exomoon the size of Earth with a surface temperature of 850 K and at a distance of five parsecs (pc) from Earth. Future instruments, such as JWST's Mid-Infrared Instrument (MIRI), have even more potential for direct imaging of exomoons. Figure 15 shows minimum temperatures and radii of exomoons detectable with MIRI at the $5\sigma$ confidence level in 10,000 seconds for a star 3 pc from the Sun (Glasse et al., 2010). Note that some exomoons shown here have temperatures and radii similar to Earth's! Thus, it is plausible that some of these hypothetical exomoons are not only directly visible with JWST, but they could potentially be habitable, in the sense of having surface temperatures that would allow liquid water to be present.

A 300 K, Earth-radius THEM is only $3 \times 10^5$ fainter at a wavelength of $\lambda \approx 14\ \mu m$ than a Sun-like star, but at the same time it can be at a large distance from it's host star. For example, at 30 AU projected separation, it would be 15″ from the star at a distance of 2 pc. At $\lambda \approx 14\ \mu m$, $\lambda/D = 0.44''$ for a D = 6.5-m telescope, which means this moon would be at $30 \times \lambda/D$. This far away from the star, the airy rings are about $3 \times 10^5$ times fainter than the core of the star and about the same intensity as the 300 K, Earth-radius moon, indicating that the detection should be possible. This example is not at the limit of JWST's sensitivity and inner working angle. The most challenging THEM detection JWST will be capable of making is a 300 K THEM as far as 4 pc from the Sun. If there is such an Earth-sized 300 K moon orbiting αCen, MIRI will be able to detect it in 8 of its 9 spectral bands with better than 15 $\sigma$ signal-to-noise in a $10^4$ s integration. Thus, directly imaging a 300 K, Earth-radius moon that is tidally heated is potentially much easier than resolving an Earth-like exoplanet orbiting in the HZ of its primary!

We can ask the question of what orbital parameters would give rise to the temperatures of moons shown in Fig. 15. The minimum detectable temperatures by MIRI for four different radii satellite ($R_s = 0.1\ R_\oplus$, 0.25 $R_\oplus$, 0.5 $R_\oplus$, and $R_\oplus$) were calculated to be $T_s = 542$ K, 355 K, 282 K, and 229 K, respectively (see Fig. 15). We can use Eq. (9) to calculate the expected orbital parameters $\beta_{ps}$ and $e_{ps}$ for these four objects. Figure 16 shows $\beta_{ps}$ and $e_{ps}$ corresponding to these four cases, indicating that a roughly Earth-sized exomoon (black line) at a distance of ten Roche radii, with an eccentricity similar to that of Titan ($e_{ps} = 0.0288$), and an effective temperature equal to the reader's room temperature could be detectable with JWST.

### 6.5 Results of exomoon searches to date

Compared to their planetary cousins, relatively few searches for exomoons have been conducted to date. The great enabler in this field has been the *Kepler* mission offering continuous, precise photometric data for years of the same targets. The downside of *Kepler* data is that the field is deep (and thus faint) and dominated by Sun-like stars, whereas K- and M-dwarfs provide a significant boost to sensitivity thanks to their smaller sizes.

Within the current literature, one can find claims that various observations are consistent with an exomoon without specifically claiming an unambiguous exomoon detection. For example, Szabo et al. (2013) suggested that many of *Kepler*'s hot-Jupiters show high TTVs that could be due to exomoons or equally other planets in the system. Others used observations of transiting exoplanets to perform serendipitous searches for exomoons, such as Brown et al. (2001) around HD 209458b, Charbonneau et al. (2006) around HD 149026b, Maciejewski et al. (2010) around WASP-3b, and Montalto et al. (2012) around WASP-3b, none of which revealed evidence for an extrasolar satellite.

At the time of writing, the only known systematic survey is the "Hunt for Exomoons with Kepler" (HEK) (Kipping et al., 2012). Using photodynamic modeling, the HEK team has surveyed seventeen Kepler planetary candidates for evidence of an extrasolar moon so far. Due to the large number of free parameters, the very complex and multimodal parameter space, the demands of full photodynamic modeling, and the need for careful Bayesian model selection, the computational requirements reported by the HEK project have been staggering compared to the a typical planet-only analysis. For example, Kepler-22b required 50 years of modern CPU time (Kipping et al., 2013b).

So far, the HEK team reports no compelling evidence for an exomoon, but they have derived strong constraints of $M_s/M_p <$ 4% for five cases (Kipping et al., 2013a; Nesvorný et al., 2012), translating into satellite masses as small as 0.07 $M_\oplus$ (Fig. 17), and $M_s < 0.5\,M_\oplus$ for the additional case of the habitable-zone planet Kepler-22b (Kipping et al., 2013b). In the latter example, an absolute mass constraint is possible thanks to the availability of radial velocities. In their latest release, Kipping et al. (2014) added another eight null detections of moons around planets transiting M dwarfs, with sensitivities to satellites as tiny as 0.36 $M_\oplus$ in the case of KOI3284.01. Despite the recent malfunction with the *Kepler* spacecraft, there is a vast volume of unanalyzed planetary candidates for moon hunting and still relatively few conducted so far. In the next one to two years then, we should see about 100 systems analyzed, which will provide a statistically meaningful constraint on the occurrence rate of large moons, $\eta_\mathbb{C}$.

## 7. Summary and Conclusions

This review highlights a remarkable new frontier of human exploration of space: the detection and characterization of moons orbiting extrasolar planets. Starting from the potentially habitable icy satellites of the Solar System (Section 2), we show that the formation of very massive satellites the mass of Mars is possible by in-situ formation in the circumplanetary disk and how a capture during binary exchange can result in a massive moon, too (Section 3). After discussing the complex orbital evolution of single and multiple moon systems, which are governed by secular perturbations and tidal evolution (Section 4), we explore illumination, tidal heating, and magnetic effects on the potential of extrasolar moons to host liquid surface water (Section 5). Finally, we explain the currently available techniques for searching for exomoons and summarize a recently initiated survey for exomoons in the data of the *Kepler* space telescope (Section 6).

From a formation point of view, habitable exomoons can exist. Mars-sized exomoons can form by either in-situ formation in the circumplanetary disks of super-Jovian planets (Section 3.1) or by gravitational capture from a former planet-moon or planet-planet binary (Section 3.2). This mass of about 0.1 $M_\oplus$ is required to let any terrestrial world be habitable by atmospheric and geological considerations (Section 5). Stellar perturbations on these moons' orbits, their tidal orbital evolution, tidal heating in the moons, and satellite-satellite interactions prevent moons in the HZs of low-mass stars from being habitable (Section 4). Moreover, while more massive giant planets should tend to produce more massive moons, strong inplanation from young, hot planets onto their moons can initiate a runaway greenhouse effect on the moons and make them at least temporarily uninhabitable (Section 5.2). With this effect becoming increasingly severe for the most massive giant planets, we identify here the formation and habitability of moons as competing processes. Although growing moons can accumulate more mass from the disks around more massive giant planets, the danger of a planet-induced runaway greenhouse effect on the moons increases, too. Finally, the host planet's intrinsic magnetic field and the stellar wind affect the moon's habitability by regulating the flux of high-energy particles (Section 5.3.2).

Exomoons in the *Kepler* data may be detectable if they belong to a class of natural satellites that does not exist in the Solar System. While the most massive known moon, Ganymede, has a mass roughly 1/40 $M_\oplus$, exomoons would need to have about ten times this mass to be traceable via their planet's TTV and TDV (Section 6.1). Another detection channel would be through a moon's direct photometric transit, which could be measurable in the *Kepler* data for moons as small as Mars or even Ganymede (Sections 6.2 and 6.3).

Combining the key predictions from formation, detection, and habitability sections, we find a favorable mass regime for the first extrasolar moons to be detected. Notably, there is a mass overlap of the most massive satellites that (i.) can possibly form in protosatellite disks around super-Jovian planets, (ii.) can be detected with current and near-future technology, and (iii.) can be habitable in terms of atmospheric stability. This regime is roughly between one and five times the mass of Mars or $0.1 - 0.5\ M_\oplus$. Heavier moons would have better odds to be habitable (Heller and Armstrong, 2014), and they would be easier to detect, though it is uncertain whether they exist.

Although the observational challenges of exomoon detection and characterization are huge, our gain in understanding planet formation and evolution would be enormous. The moons of Earth, Jupiter, Saturn, and Neptune have proven to be fundamentally important tracers of the formation of the Solar System – most notably of the formation of Earth and life – and so the characterization of exomoon systems engenders the power of probing the origin of individual extrasolar planets (Withers and Barnes, 2010). Just as the different architectures of the Jovian and Saturnian satellite systems are records of Jupiter's gap opening in the early gas disk around the Sun, of an inner cavity in Jupiter's very own circumplanetary disk, and of the evolution of the $H_2O$ snow line in the disks, the architectures of exomoon systems could trace their planets' histories, too. We conclude that any near-future detection of an exomoon, be it in the *Kepler* data or by observations with similar accuracy, could fundamentally challenge formation and evolution theories of planets and satellites.

Another avail of exomoon detections could lie in a drastic increase of potentially habitable worlds. With most known planets in the stellar HZ being gas giants between the sizes of Neptune and Jupiter rather than terrestrial planets, the moons of giant planets could actually be the most numerous population of habitable worlds.

## 8. Outlook

On the theoretical front, improvements of our understanding of planet and satellite formation might help to focus exomoon searches on the most promising host planets. The formation and movement of water ice lines in the disks around young giant planets, as an example, has a major effect on the total mass of solids available for moon formation. Preliminary studies show that super-Jovian planets might have formed water-rich giant moons the size of Mars or even larger (Heller and Pudritz, in prep.), in agreement with the scaling relation suggested by Canup and Ward (2006). Hence, even future null detections of moons around the biggest planets could help to assess the conditions in circumplanetary disks. As an example, the absence of giant moons could indicate hotter planetary environments, in which the formation of ices is prevented, than are currently assumed. Earth-mass moons, however, are very hard to form with any of the viable formation theories, and consequently the suggested low abundance or absence (Kipping et al., 2014) of such moons may not help to constrain models immediately. However, in the most favorable conditions, where a somewhat super-Jovian-mass planet with an entourage of an oversized Galilean-style moon system transits a photometrically quiet M or K dwarf star, a handful of detections might be feasible with the available *Kepler* data (Kipping et al., 2012; Heller 2014) and therefore might confirm the propriety of moon formation theories for extrasolar planets.

Another largely unexplored aspect of habitable moon formation concerns the capture and orbital stability of moons during planet migration. A few dozen giant planets near, or in, their host stars' HZs have been detected by radial velocity techniques (see Fig. 1 in Heller and Barnes, 2014), most of which cannot possibly have formed at their current orbital locations, as giant planets' cores are assumed to form beyond the circumstellar snow lines (Pollack et al., 1996; Ward, 1997). This prompts the compelling question of whether migrating giant planets, which end up in their host star's HZ, can capture terrestrial planets into stable satellite orbits during migration. A range of studies have addressed the post-capture evolution or stability of moons (Barnes and O'Brien, 2002; Domingo et al., 2006; Donnison et al., 2006; Porter and Grundy, 2011; Sasaki et al., 2012; Williams, 2013) and the loss of moons during planet migration towards the hot Jupiter regime (Namouni, 2010). But dynamical simulations of the capture scenarios around migrating giant planets near the HZ are still required in order to guide searches for habitable exomoons.

NASA's all-sky survey with the *Transiting Exoplanet Survey Satellite* (*TESS*), to be launched in 2017, will use the transit method to search for planets orbiting bright stars with orbital periods ≤ 72 days (Deming et al., 2009). As stellar perturbations play a key role in the orbital stability of moons orbiting giant planets (see Section 4), *TESS* will hardly find such systems. The proposed ESA space mission *PLAnetary Transits and Oscillations of stars* (*PLATO 2.0*) (Rauer et al., 2013), with launch envisioned for 2022-2024, however, has been shown capable of finding exomoons as small as $0.5\ R_\oplus$, that is, the size of Mars (Simon et al., 2012; Heller 2014). Since the target stars will be mostly K and M dwarfs in the solar neighborhood, precise radius and mass constraints on potentially discovered exomoons will be possible (Kipping, 2010). What is more, *PLATO*'s long pointings of two to three years would allow for exomoon detections in the HZs of K and M dwarfs.

Future exomoon surveys may benefit by expanding their field of view while keeping a continuous staring mode and using the redder part of the spectrum to look at K and M dwarfs. Targeting these low-mass stars increases the relative dip in the light curves due to an exomoon as well as the transit frequency, and it decreases the minimum transit duration of a planet capable of hosting long-lived stable moons. Searches for exomoons around specific transiting planets will need to target relatively inactive stars or stars with noise spectra that are amenable to filtering (Carter and Winn, 2009) in order to reduce the effect of starspot modulation and errors due to spot crossing events. While recently developed procedures of white and red noise filtering in stellar light curves represent major steps towards secure moon detections, they do not directly address additional timing noise due to spots on the stellar limb (Silva-Válio, 2010; Barros et al., 2013) and the empirical correlation between the timing error and the slope of the out of transit light curve (Mazeh et al., 2013). Further progress, either through the modeling-based approach used by the HEK project, the filtering approach suggested by Carter and Winn (2009), or the observational approach used by Mazeh et al. (2013) is required to securely detect and constrain moons of *Kepler* planets. Finally, telescopes capable of multi-color measurements will help discriminate between starspot occupations on the one hand and planet/moon transits on the other hand.

*JWST* will be sensitive to a hypothetical family of tidally heated exomoons (THEMs) at distances as far as 4 pc from the Sun and as a cool as 300 K. These moons would need to be roughly Earth-sized and, together with the host planets, separated from their host star by at least 0.5″. Future surveys that aim to detect smaller and colder THEMs would need to operate in the mid- to far-infrared with sub-microjansky sensitivity. Although THEMs need not be close to their host star to be detectable, instruments able to probe smaller inner working angles are likely to detect more THEMs simply because there will presumably be more planets closer into the star that can host THEMs. A preliminary imaging search for THEM around the nearest stars with the use of archival Spitzer IRAC data has recently been completed and will deliver the first constraints on the occurrence of THEMs (Limbach and Turner in prep).

Once an exomoon has been detected around a giant planet in the stellar HZ, scientists will try to discern whether it is actually inhabited. Kaltenegger (2010) showed that *JWST* would be able to spectroscopically characterize the atmospheres of transiting exomoons in nearby M dwarf systems, provided their circumplanetary orbits are wide enough to allow for separate transits. In her simulations, the spectroscopic signatures of the biologically relevant molecules $H_2O$, $CO_2$, and $O_3$ could be observable for HZ exomoons transiting M5 to M9 stars as far as 10 pc from the Sun. Although stellar perturbations on those moon's orbits could force the latter to be eccentric and thereby generate substantial tidal heating (Heller, 2012), the threat of a runaway greenhouse effect would be weak for moons in wide orbits because tidal heating scales inversely to a high power in planet-moon distance.

Exomoon science will benefit from results of the *Jupiter Icy Moons Explorer* (*JUICE*) (*JUICE* Assessment Study Report[28], 2011; Grasset et al., 2013), the first space mission designated to explore the emergence of habitable worlds around giant planets. Scheduled to launch in 2022 and to arrive at Jupiter in 2030, *JUICE* will constrain tidal processes and gravitational interaction in the Galilean system and thereby help to calibrate secular-tidal models for the orbital evolution of moons around extrasolar planets. As an example, the tidal response of Ganymede's icy shell will be measured by laser altimetry and radio science experiments. The amplitudes of periodic surface deformations are suspected to be about 7 to 8 m in case of a shallow subsurface ocean, but only some 10 cm if the ocean is deeper than roughly 100 km or not present at all. With Ganymede being one of the three solid bodies in the Solar System – besides Mercury and Earth – that currently generate a magnetic dipole field (Kivelson et al., 2002), *JUICE* could constrain models for the generation of intrinsic magnetic fields on extrasolar moons, which could be crucial for their habitability (Baumstark-Khan and Facius, 2002; Heller and Zuluaga, 2013). Measuring the bodily and structural properties of the Galilean moons, the mission will also constrain formation models for moon systems in general.

In view of the unanticipated discoveries of planets around pulsars (Wolszczan and Frail, 1992), Jupiter-mass planets in orbits extremely close to their stars (Mayor and Queloz, 1995), planets orbiting binary stars (Doyle et al., 2011), and small-scale planetary systems that resemble the satellite system of Jupiter (Muirhead et al., 2012), the discovery of the first exomoon beckons, and promises yet another revolution in our understanding of the universe.

## Acknowledgements

The helpful comments of two referees are very much appreciated. We thank Alexis Carlotti, Jill Knapp, Matt Mountain, George Rieke, Dave Spiegel and Scott Tremaine for useful conversations and Ted Stryk for granting permission to use a reprocessed image of Europa. René Heller is supported by the Origins Institute at McMaster University and by the Canadian Astrobiology Training Program, a Collaborative Research and Training Experience Program funded by the Natural Sciences and Engineering Research Council of Canada (NSERC). Darren Williams is a member of the Center for Exoplanets and Habitable Worlds, which is supported by the Pennsylvania State University, the Eberly College of Science, and the Pennsylvania Space Grant Consortium. Takanori Sasaki was supported by a grant for the Global COE Program, "From the Earth to 'Earths'", MEXT, Japan, and Grant-in-Aid for Young Scientists (B), JSPS KAKENHI Grant Number 24740120. Rory Barnes acknowledges support from NSF grant AST-1108882 and the NASA Astrobiology Institute's Virtual Planetary Laboratory lead team under cooperative agreement no. NNH05ZDA001C. Jorge I. Zuluaga is supported by CODI/UdeA. This research has been supported in part by World Premier International Research Center Initiative, MEXT, Japan. This work has made use of NASA's Astrophysics Data System Bibliographic Services.

---

[28] Yellow Book available at http://sci.esa.int/juice